\documentclass{ws-ijmpe}
\usepackage[super,compress]{cite}
\begin{document}

\markboth{W. Weise}{From QCD symmetries to nuclei and neutron stars}

\catchline{}{}{}{}{}

\title{From QCD Symmetries to Nuclei and Neutron Stars\footnote{Dedicated in memory of Ernest M. Henley (1924 - 2017).}}

\author{Wolfram Weise}

\address{Physics Department, Technical University of Munich, 85748 Garching, Germany
weise@tum.de}

\maketitle

\begin{history}
\received{Day Month Year}
\revised{Day Month Year}
\end{history}

\begin{abstract}
Global symmetries and symmetry breaking patterns of QCD with light quarks, in particular chiral symmetry, provide basic guidance not only for low-energy hadron physics but also for nuclear forces and the nuclear many-body problem. Recent developments of Chiral Effective Field Theory applications to nuclear and neutron matter are summarized, with special emphasis on a (non-perturbative) extension using functional renormalisation group methods. Topics include: nuclear thermodynamics, extrapolations to dense baryonic matter and constraints from neutron star observables. 
\end{abstract}

\keywords{Low-energy QCD; chiral symmetry; dense nuclear matter; neutron stars.}

\ccode{PACS numbers: 12.39.Fe,97.60.Jd,26.60.Kp,26.60.-c}


\section{Introductory Survey: \\QCD Symmetry Breaking Patterns and Scales}
Understanding nuclei and dense baryonic matter at the interface with quantum chromodynamics (QCD), the underlying theory of the strong interaction, is one of the pending challenges in nuclear many-body theory.
Global symmetries of QCD provide guidance for constructing effective field theories that represent low-energy QCD and thus establish a conceptual frame for nuclear physics. In its sector with the two lightest ($u$ and $d$) quarks, two such basic symmetries of QCD are a key to understanding low-energy energy hadron structure as well as nuclear forces: {\it scale invariance} and {\it chiral symmetry}. Given the small quark masses\cite{PDG}, $m_u\simeq 2$ MeV and $m_d\simeq 5$ MeV (at a renormalization scale $\mu\simeq 2$ GeV), a useful starting point is QCD with a massless isospin doublet $\psi=(u,d)^\top$ of quarks. In this limit, scale and chiral invariance are {\it exact} symmetries of QCD.

\subsection{Scale invariance, trace anomaly and the mass of the nucleon}

QCD with massless quarks has no dimensional parameter. A scale transformation of the quark and gluon fields, $\psi(x)\rightarrow \lambda^{3/2}\psi(\lambda x)$ and $A^\mu(x) \rightarrow\lambda A^\mu(\lambda x)$ with an arbitrary constant $\lambda$, leaves the QCD action $S = \int d^4x\, {\cal L}_{QCD}$ invariant. The associated conserved current, $S^\mu = x_\nu\Theta^{\mu\nu}$, involves the energy-momentum tensor $\Theta^{\mu\nu}$ derived from the QCD Lagrangian ${\cal L}_{QCD}$. At the classical level this current is conserved, i.e. $\partial_\mu S^\mu = \Theta_\mu^\mu = 0$. QCD as a quantum field theory, on the other hand, is subject to renormalization which introduces a scale, $\Lambda_{QCD}$, so that scale invariance is broken. Consequently, the trace of the QCD energy-momentum tensor is non-vanishing. It is given by the {\it trace anomaly}: 
\begin{equation}
\Theta_\mu^\mu = {\beta(g)\over g}\,Tr\left[G_{\mu\nu}G^{\mu\nu}\right] ~,
\end{equation}
involving the gluon field tensor $G_a^{\mu\nu}$ (with the SU(3) color index $a=1,\dots,8$) and $Tr\left[G_{\mu\nu}G^{\mu\nu}\right] = {1\over 2}\sum_a G_{\mu\nu,a}G^{\mu\nu}_a = \boldsymbol{B}^2 -\boldsymbol{E}^2$ in terms of color-magnetic and color-electric fields. The leading-order QCD beta function is $\beta(g) = -{\beta_0\over 16\pi^2}g^3$ with $\beta_0 = 11-{2\over 3}N_f = 9$ for $N_f = 3$ quark flavors, and we write $\alpha_s = {g^2\over 4\pi}$ as usual.

A prominent phenomenon in QCD is the emergence of the nucleon mass, $M_N$, as a characteristic 1 GeV scale starting from (almost) massless $u$ and $d$ quarks. Gluon dynamics and the trace anomaly are at the origin of this mass generation mechanism which is thus responsible for almost all of the visible mass in the universe. 

Consider the matrix element of the energy-momentum tensor,
\begin{equation}
\langle N(P)|\Theta^{\mu\nu}|N(P)\rangle = (P^\mu P^\nu/M_N)\,\bar{U}(P)\,U(P)~, \nonumber
\end{equation}
of a nucleon with four-momentum $P^\mu$. The Dirac spinor $U(P)$ of the nucleon is normalized as $\bar{U}\,U = 2M_N$. The nucleon mass, $M_N = \sqrt{P^\mu P_\mu}$, is determined by taking the trace of the energy-momentum tensor. For massless quarks we have:  
\begin{eqnarray}
\langle N(P)|\Theta_\mu^\mu|N(P)\rangle &=& \langle N(P)|-{9\alpha_s\over 4\pi} Tr\left[G_{\mu\nu}G^{\mu\nu}\right]|N(P)\rangle\nonumber\\
 &=&  {9\over 4}\langle N(P)|{\alpha_s\over \pi} (\boldsymbol{E}^2- \boldsymbol{B}^2)|N(P)\rangle = M_N^{(0)}\, \bar{U}(P)\,U(P)~.
\end{eqnarray}
The resulting mass\cite{Ji1995}, $M_N^{(0)}$, differs by about $10 \% $ from the physical nucleon mass, underlining its prominent gluon-dynamical origin.
Nonzero quark masses with $m_q \equiv {1\over 2}(m_u + m_d)$ add a comparatively small piece (the sigma term) $\sigma_N = \langle N|m_q(\bar{u}u+\bar{d}d)|N\rangle /(2M_N)$, such that the physical nucleon mass becomes
\begin{equation}
M_N = M_N^{(0)} + \sigma_N+ \sigma_s~,
\end{equation}
where $\sigma_s\propto \langle N|m_s\,\bar{s}s|N\rangle$ denotes the additional contribution from the strange-quark mass.
The empirical value of the nucleon sigma term, deduced from low-energy pion-nucleon scattering data, has for a long time been in the range\cite{Gasser1991} $\sigma_N \simeq (45\pm 8)$ MeV. Recent advanced analysis\cite{Ruiz2018} arrives at a significantly larger value: $\sigma_N \simeq (59\pm 5)$ MeV, while lattice QCD computations\cite{Duerr2016} suggest a smaller value (38 MeV with about $20\%$ uncertainty). The apparent tension between these numbers is still an issue. The strangeness sigma term of the nucleon, $\sigma_s$, is likewise uncertain.

\subsection{Spontaneous chiral symmetry breaking and the pion}

In the limit of vanishing quark masses, $m_u = m_d = 0$, QCD has an exact $SU(2)_L\times SU(2)_R$ chiral symmetry: in absence of quark mass terms the QCD Lagrangian is invariant under separate isospin-$SU(2)$ rotations of the left- and right-handed quark doublets, $\psi_L = (u_L,d_L)^\top$ and $\psi_R = (u_R,d_R)^\top$. The associated conserved currents of the left- and right-handed quarks, $J^\mu_{L,R} = \bar{\psi}_{L,R}\gamma^\mu\psi_{L,R}$, are conveniently rewritten as isovector axial and vector currents, 
\begin{eqnarray}
\boldsymbol{A}^\mu(x) = \bar{\psi}(x)\,\gamma^\mu\gamma_5\,{\boldsymbol{\tau}\over 2}\,\psi(x)~, ~~~\boldsymbol{V}^\mu(x) = \bar{\psi}(x)\,\gamma^\mu\,{\boldsymbol{\tau}\over 2}\,\psi(x)~,
\end{eqnarray}
with $\partial_\mu\boldsymbol{A}^\mu=\partial_\mu\boldsymbol{V}^\mu=0$. Chiral symmetry remains intact in its trivial (Wigner-Weyl) realisation at any order of perturbation theory in high-energy QCD. However, the non-perturbative nature of strong interactions at low energy induces a qualitatively different scenario: the Nambu-Goldstone realisation of spontaneously broken chiral symmetry. 

The spontaneous breaking of a global symmetry implies the existence of massless Nambu-Goldstone (NG) bosons. In the present case, the original chiral $SU(2)_L\times SU(2)_R$ symmetry of QCD with massless quarks is reduced to isospin $SU(2)_V$ which remains a good symmetry, together with $U(1)_B$ for baryon number. The NG bosons are identified with the isospin triplet of pions, $\boldsymbol{\pi} = (\pi_1,\pi_2,\pi_3)$. At the same time spontaneous chiral symmetry breaking implies a qualitative change of the QCD vacuum, generating a scalar condensate of quark-antiquark pairs, the quark condensate $\langle \bar{q}q\rangle = \langle\bar{u}u + \bar{d}d\rangle$. Historically, the mechanism of spontaneous chiral symmetry breaking with the emergence of pions as NG bosons and a non-zero quark condensate was elegantly displayed by the Nambu \& Jona-Lasinio (NJL) model\cite{NJL1961} which subsequently enjoyed various extensions and multiple applications\cite{VW1991,HK1994} in hadron physics.

Explicit symmetry breaking by the non-zero quark masses shifts the pion mass from zero to its physical value. The pion mass is connected to the bare quark mass, $m_q = {1\over 2}(m_u + m_d)$, through the Gell-Mann - Oakes - Renner relation\cite{GOR1968} derived originally from current algebra and PCAC:
\begin{equation}
m_\pi^2\,f_\pi^2 = -m_q\,\langle\bar{q}q\rangle + {\mathcal O}(m_q^2,\, m_q^2 \log m_q)~.
\end{equation}
It involves the pion decay constant, $f_\pi$, defined by the matrix element connecting the pion with the QCD vacuum via the isovector axial current:
\begin{equation}
\langle 0|A^\mu_i(0)|\pi_j(q)\rangle = i\delta_{ij}\,q^\mu f_\pi~.
\end{equation}
The empirical pion decay constant, $f_\pi \simeq 92.2$ MeV, is slightly larger than its value in the chiral limit, $f_\pi^{(0)} \simeq 86$ MeV. Like the quark condensate $\langle\bar{q}q\rangle$, this decay constant is a measure of spontaneously broken chiral symmetry with a characteristic scale, $\Lambda_\chi \sim 4\pi\,f_\pi^{(0)}\simeq 1$ GeV. The non-zero pion mass, $m_\pi \ll \Lambda_\chi$, is a reflection of the explicit chiral symmetry breaking by the small  $u$ and $d$ quark masses, with $m^2_{\pi}$ proportional to $m_q$. The following correspondence holds for a comparison of these ``small" scales in terms of quark and hadron language:
\begin{equation}
{m_q\over \Lambda_{QCD}}\sim \left({m_\pi\over 4\pi f_\pi}\right)^2 \simeq 1.4\cdot 10^{-2}~,
\end{equation}
using $\Lambda_{QCD}\sim 0.25$ GeV. 

Unlike the pion mass and decay constant, the quark masses $m_{u,d}$ as well as the quark condensate $\langle 0|\bar{q} q |0\rangle$ are scale-dependent quantities and therefore not separately observable. Only their product is renormalization group invariant. At a renormalization scale $\mu\simeq 2$ GeV, a typical average quark mass $m_q \simeq 3.5$ MeV goes together with a condensate $\langle\bar{q} q\rangle = \simeq -(0.36\,\textrm{GeV})^3 \simeq -6\,\textrm{fm}^{-3}$.

\subsection{Nucleons and pions in low-energy QCD}

Consider now how the nucleon is to be viewed under the aspect of spontaneously broken chiral symmetry. As a starting point imagine three massless quarks forming a color-singlet with one unit of baryon number and total spin $1/2$. Embedded in the non-trivial QCD vacuum, this originally massless trio of quarks acts as a seed for the formation of the nucleon with its large mass $M_N^{(0)}$ as just outlined. The massive nucleon emerges as a complex system in which the three valence quarks, accompanied by a strong gluonic field and a sea of quark-antiquark pairs, are localized in a small volume with a radius of less than a Fermi. 

Localization (confinement) of the valence quarks in a finite volume implies breaking of chiral symmetry. The quark helicity, $h = \vec{\sigma}\cdot\vec{p}/|\vec{p}\,|$, changes sign when the quark momentum $\vec{p}$ is reflected at the confining boundary whereas the spin $\vec{\sigma}$ keeps its direction: left- and right-handed quark chiralities mix. This symmetry breaking mechanism has its origin again in the strong-coupling dynamics of low-energy QCD. 

The following low-energy QCD based picture relevant to nuclear physics thus emerges\cite{BR1979,TTM1981,TW2001}. The nucleon mass is predominantly generated by gluon dynamics through the trace anomaly. The same underlying strong-interaction processes induce the spontaneous breaking of chiral symmetry. The pion is singled out as an almost massles Nambu-Goldstone boson, a collective quark-antiquark mode of the non-perturbative QCD vacuum. The three valence quarks in the nucleon, carrying one unit of baryon number, are localized in a small spatial volume. Strong vacuum polarization effects surround this baryonic core with multiple quark-antiquark pairs. The low-energy physics associated with this mesonic ($\bar{q}q$) cloud is governed by pions as NG bosons of spontaneously broken chiral symmetry, in such a way that the total axial vector current of the ``core + cloud" system is conserved apart from small mass corrections (PCAC). Interactions between two nucleons at long and intermediate distances are generated as an inward-bound hierarchy of one-, two-, multi-pion exchange processes. The framework for systematically organizing this hierarchy is Chiral Effective Field Theory (ChEFT). It is based on the separation of scales that delineates low energies and momenta characteristic of nuclear physics from the chiral symmetry breaking ``gap" scale, $\Lambda_\chi = 4\pi f_\pi \sim 1$ GeV $\sim M_N$.

\begin{figure}[th]
\centerline{\includegraphics[width=6.5cm]{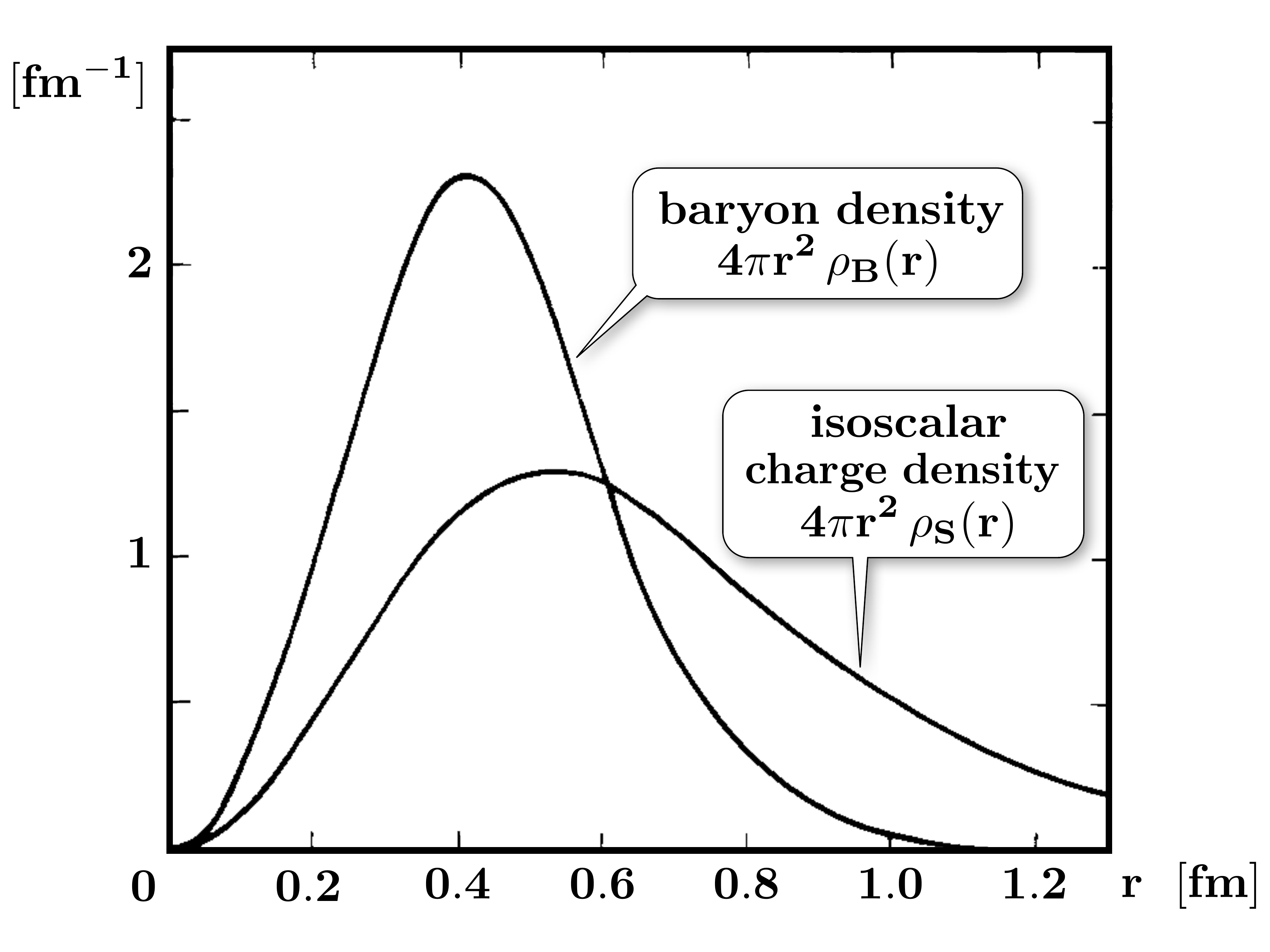}}
\caption{Distributions (multiplied by $4\pi r^2$) of baryon number, $\rho_B(r)$, and isoscalar electric charge, $\rho_S(r)= \rho({proton}) + \rho({neutron})$, resulting from a chiral soliton model of the nucleon \cite{KMW1987} .}
\label{fig:1}
\end{figure}

In this context it is instructive to examine the typical size scales of the nucleon itself. As mentioned the ``chiral" nucleon consists of a compact valence quark core surrounded by a multi-pion cloud. Early concepts of the nucleon as a topological soliton derived from a non-linear chiral meson Lagrangian have drawn a picture\cite{KMW1987} of the distributions of baryon number, $\rho_B(r)$, and of (isoscalar) electric charge, $\rho_S(r) \equiv\rho_{p}(r) + \rho_{n}(r)$, as shown in Fig.\,\ref{fig:1}. The root-mean-square radius of the baryonic core turns out to be $\langle r_B^2\rangle^{1/2} \simeq 0.5$ fm. The isoscalar charge radius of the nucleon, determined primarily by its charged meson cloud, is significantly larger: $\langle r_S^2\rangle^{1/2}\simeq 0.8$ fm. The ratio of the corresponding volumes, ${\mathcal V}(baryon~no.) / {\mathcal V}(charge) \sim 0.2$, indicates yet another separation of scales relevant to the window of applicability for ChEFT.

The detailed investigation and analysis of deeply virtual Compton scattering measurements\cite{CLAS2015a,CLAS2015b} at J-Lab can provide further empirical information on these scales. Such data are expected to map out the transverse profile of the proton at different ranges of longitudinal quark momentum (i.e.  Bjorken-$x$), thus interpolating between sections dominated by valence quarks $(0.3\lesssim x_B \lesssim 0.5)$ and those prominently involving the sea of quark-antiquark pairs $(x_B < 0.2)$. Recent and ongoing studies combined with a Regge picture suggest an $x_B$-dependence of the transverse proton size as follows\cite{DGV2017}: 
\begin{equation}
\langle \boldsymbol{b}^2\rangle \simeq 0.16\,\text{fm}^2\,\cdot\ln\left({1\over x_B}\right)~.
\end{equation}
In the valence quark region, ${1\over 3} < x_B < {1\over 2}$, this implies a ``core" size 
\begin{equation}
\text{R}_{core} = \sqrt{{3\over 2} \langle \boldsymbol{b}^2\rangle} \simeq 0.4 - 0.5~\text{fm}~, 
\end{equation}
consistent with the previously discussed radius of the baryon density in the chiral soliton model of the nucleon. 

\subsection{Towards compressed baryonic matter}

\begin{figure}[th]
\centerline{\includegraphics[width=9.5cm]{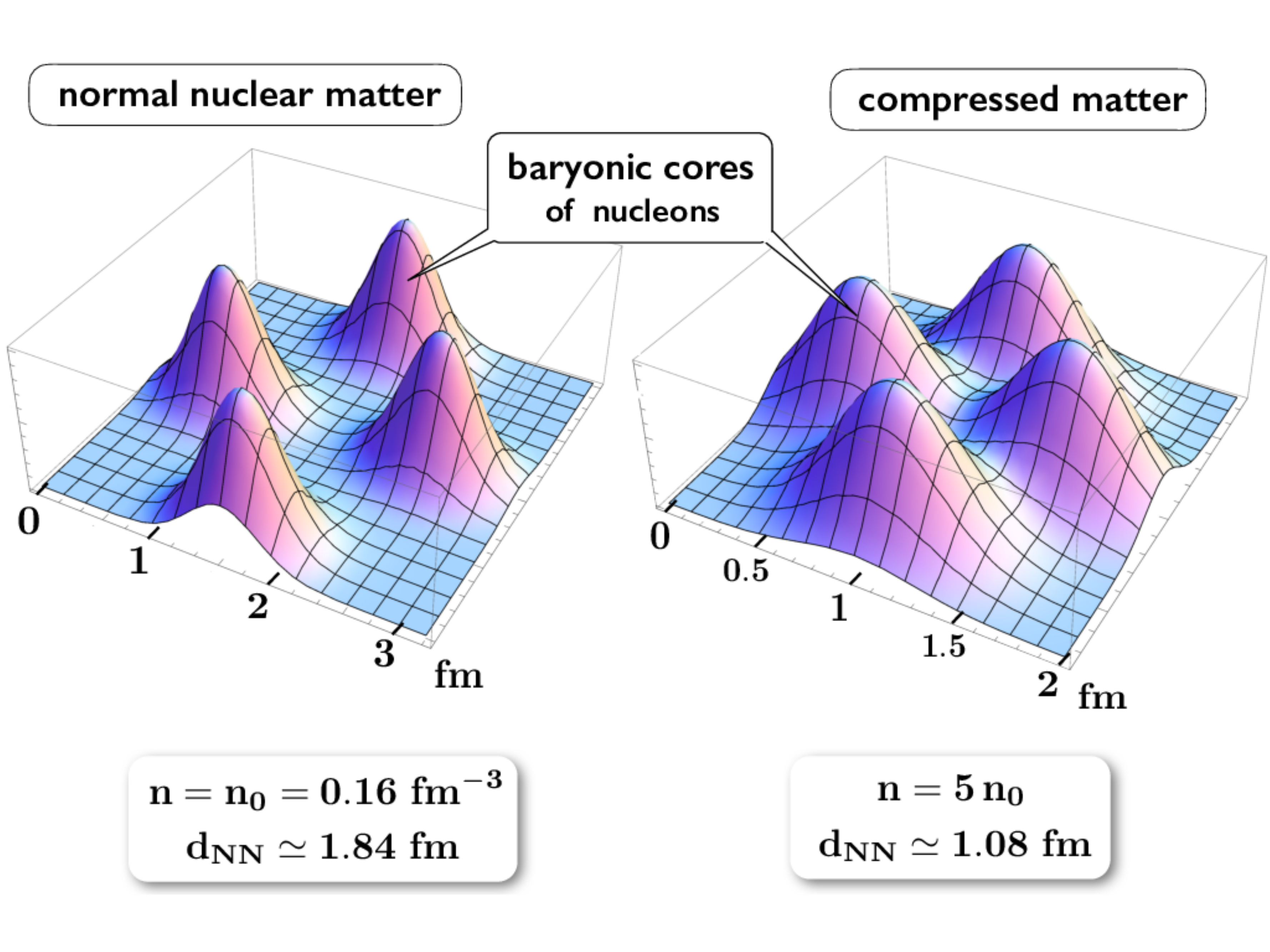}}
\caption{Schematic picture of baryonic matter at normal nuclear density ($n_0 = 0.16$ fm$^{-3}$, left) and at $n = 5\,n_0$ (right). The individual baryon number density distributions are Gaussians with an r.m.s. radius $\sqrt{\langle r_B^2\rangle}=0.5$ fm.}
\label{fig:2}
\end{figure}

Assuming a typical 1/2 Fermi radius of the baryon core, let us consider compressed baryonic matter and examine up to which baryon densities, $n = B/\cal{V}$, one can still expect nucleons (rather than free-floating quarks) to be the relevant baryonic degrees of freedom. A schematic picture is drawn in Fig.\,\ref{fig:2}, illustrating a piece of baryonic matter as a set of Gaussian distributions. Each of these Gaussians approximates very well the baryon density of the nucleon as a chiral soliton, Fig.\,\ref{fig:1}. At the density of normal nuclear matter, $n_0 = 0.16$ fm$^{-3}$, the baryonic cores are well separated by an average distance $d_{NN} \sim n_0^{-1/3}\simeq 1.8$ fm. Pions couple to these baryonic sources and act in the space between these cores. The pion field incorporates multiple exchanges of pions between nucleons, and those mechanisms are properly dealt with in chiral EFT.

 While these longer-range pionic field configurations will undergo drastic changes in highly compressed baryonic matter, the sizes of the compact baryonic cores themselves are expected to remain stable as long as they continue to be separated. As illustrated in Fig.\,\ref{fig:2}, even at $n = 5\,n_0$, corresponding to an average distance $d_{NN} \simeq 1.1$ fm between nucleons, the individual baryon distributions are indeed still well identifiable with just small overlaps at their touching surfaces. At such high densities the (non-linear) pion field between baryonic sources is accumulating much strength. Non-perturbative chiral field theory methods must be employed to treat these strong-field configurations.

In this picture the nucleons loose their identities once the density reaches $n\gtrsim 10\,n_0$ and the baryon distributions begin to merge (percolate). Quark matter with strong pairing (diquarks) is supposed to take over\cite{Baym2018} . However, such extreme densities are already beyond the baryon densities typically encountered in the central regions of neutron stars if their radii are larger than about ten kilometers. 
One should also note that the strong short-distance repulsion between nucleons tends to make a ``soft" merging scenario for nucleons energetically very expensive. The repulsive hard core of the NN interaction with its typical range of about half a Fermi has a long phenomenological history in nuclear physics. More recently this hard core has been established by deducing an equivalent local NN potential from lattice QCD computations\cite{AHI2010,Aoki2013} .

\section{Chiral Effective Field Theory and Related Approaches to Nuclear and Neutron Matter}

Chiral effective field theory\cite{Weinberg1996} starts out as a non-linear theory of Nambu-Goldstone pions and their interactions, with symmetry-breaking mass terms added. Nucleons are introduced as ``heavy" sources of the NG bosons. A systematically organized low-energy expansion in derivatives of the pion field (power-counting) provides a remarkably successful quantitative description (chiral perturbation theory\cite{GL1984}) of pion-pion scattering and of pion-nucleon interactions at momenta and energies small compared to the symmetry breaking scale, $\Lambda_\chi\sim 4\pi f_\pi$. ChEFT with inclusion of nucleons\cite{BKM1995} is also the basis for a highly successful theory of the nucleon-nucleon interaction. In recent years this has become the widely accepted input for the treatment of nuclear many-body problems\footnote{See e.g. references\cite{HKW2013,HRW2016} for recent overviews.}. 

\subsection{Chiral nuclear forces}

\begin{figure}
\centerline{\includegraphics[width=7cm] {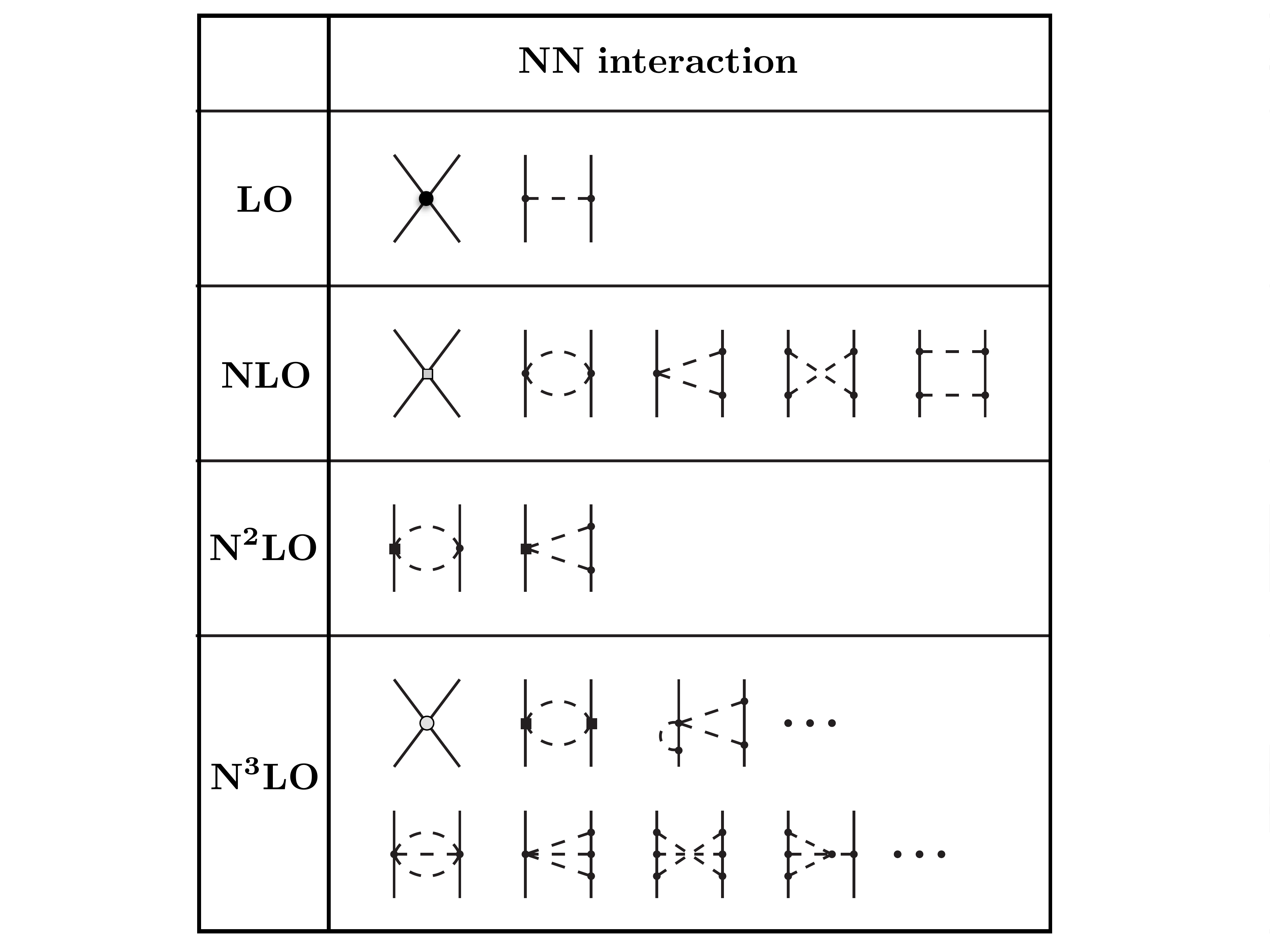}}
\caption{Diagrammatic hierarchy of chiral nuclear forces (single and multiple pion exchange) plus contact terms up to fourth order in powers of $Q/\Lambda$.}
\label{fig:3}
\end{figure}

Nuclear forces in ChEFT are constructed in terms of explicit one- and
multi-pion exchange processes, constrained by chiral symmetry, plus a complete set of contact terms
encoding unresolved short-distance dynamics \cite{evgenireview,hammerreview,machleidtreview}. A systematic hierarchy of diagrammatic contributions (see Fig.\,\ref{fig:3}) is organized in powers of $Q/\Lambda$, where $Q$ stands generically for small momenta or the pion mass, and $\Lambda < \Lambda_\chi = 4\pi f_\pi \sim 1$\,GeV is a conveniently chosen cutoff, intermediate with reference to the chiral symmetry breaking scale. Two-body forces at leading-order (LO), corresponding to $(Q/\Lambda)^0$, include the one-pion exchange contribution together with two contact terms acting in relative $S$-waves.
Next-to-leading order (NLO) terms proportional to $(Q/\Lambda)^2$ include two-pion exchange diagrams
involving vertices of the chiral $\pi N$-Lagrangian. These NLO contributions generate isovector central, isoscalar
spin-spin and isoscalar tensor forces. Detailed expressions are found in the mentioned literature. Additional polynomial pieces produced
by pion loops can be absorbed into the set of contact terms at NLO which features seven low-energy constants associated with all combinations of spin, isospin and momentum operators appearing at that order. These low-energy constants are adjusted to fit NN scattering phase shifts.

At order N$^2$LO the most important pieces from chiral two-pion exchange arise, those that generate the prominent intermediate-range attraction in the
isoscalar central NN channel and reduce the overly strong one-pion-exchange isovector tensor force.
These terms \cite{KBW1997} involve chiral $\pi\pi NN$ contact couplings. They include the prominent effects of $\Delta(1232)$-isobar degrees of freedom in intermediate steps of the two-pion exchange process\footnote{In fact, treating the $\Delta(1232)$ as an additional explicit baryonic degree of freedom promotes these mechanisms from N$^2$LO to NLO.}. In addition there are relativistic $1/M_N$ corrections to $2\pi$-exchange \cite{KBW1997,evgenireview}. 

Chiral NN potentials constructed up to order N$^3$LO (i.e., fourth order in $Q/\Lambda$) include two-loop $2\pi$-exchange processes, $3\pi$-exchange terms plus contact forces quartic in the momenta and parameterized by 15 additional low-energy constants. When solving the Lippmann-Schwinger equation the potential is multiplied by an exponential regulator function with a cutoff scale $\Lambda = 400 - 700\,$MeV $<\Lambda_\chi$ in order to
restrict integrations to the low-momentum region where chiral effective field theory is applicable.
At order N$^3$LO the chiral NN interaction reaches the quality of a ``high-precision" ($\chi^2/d.o.f.\sim 1$) potential in reproducing empirical NN scattering
phase shifts and deuteron properties \cite{hammerreview,CHIMS2013}. At the same time it provides the foundation for systematic nuclear structure studies of few- and many-body systems. Current state-of-the-art potentials\cite{MS2016}  have progressed to chiral order N$^4$LO (fifth order in $Q/\Lambda$) and achieved still further substantial improvements in comparison with empirical NN two-body data, including peripheral phase shifts and $np$ polarization observables \cite{EKMN2015, EKM2015} . Convergence has been successfully demonstrated by investigating dominant contributions of order N$^5$LO \cite{EKMN2015b} . 

\begin{figure}
\centerline{\includegraphics[width=5cm] {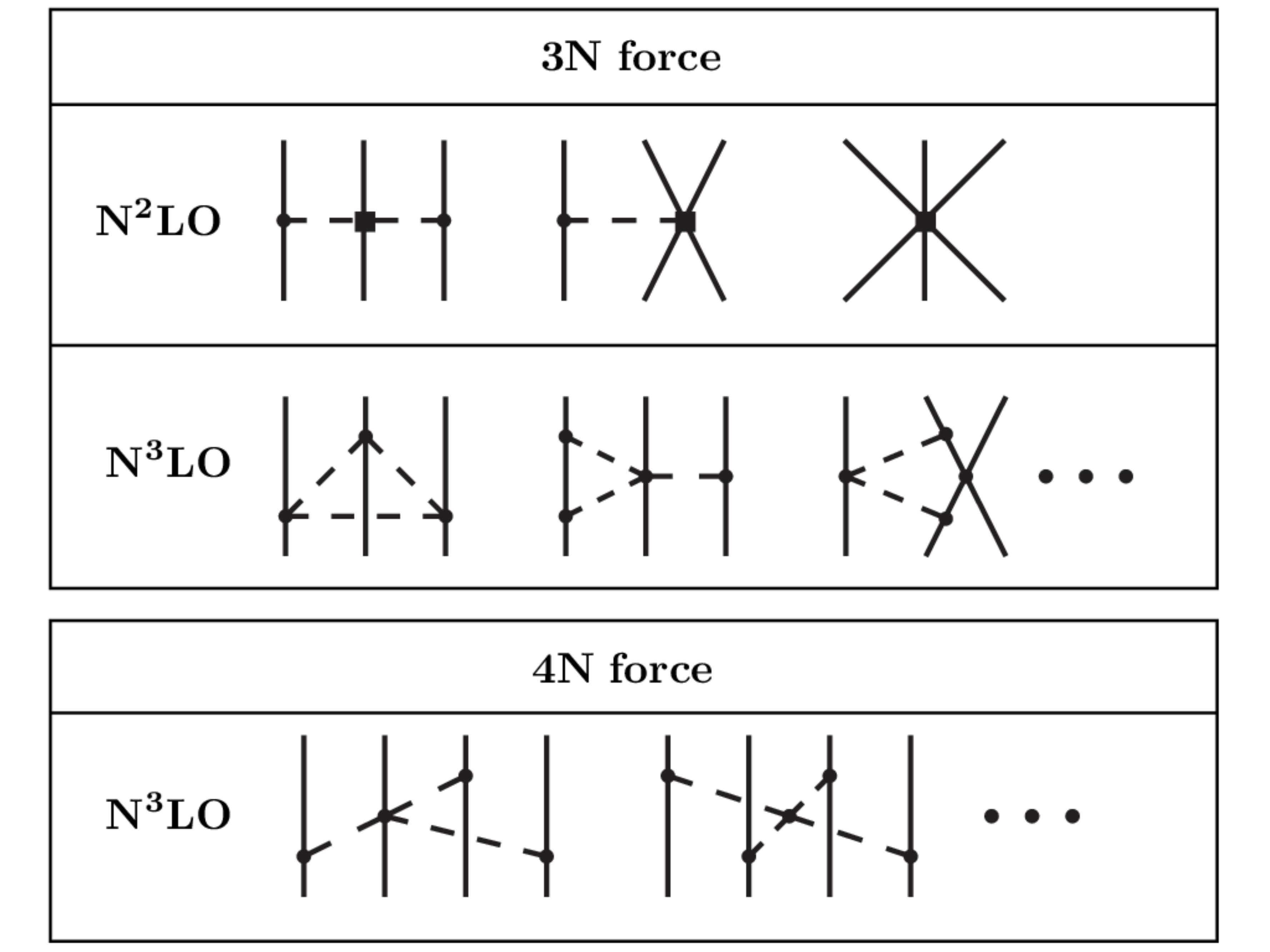}}
\caption{Three- and four-body nuclear forces generated in chiral EFT at orders N$^2$LO and N$^3$LO.}
\label{fig:4}
\end{figure}

Recent years have seen many applications of chiral two- and three-nucleon interactions in various nuclear structure computations and calculations of nuclear and neutron matter (see, e.g., \cite{HNS2013,LENPIC2016,Coraggio2016} and refs.\,therein). The strength of the ChEFT approach is its well-defined perturbative hierarchy, increasing order by order in the complexity of two- and many-body interactions. Three-body forces first enter at order N$^2$LO as illustrated in Fig.\,\ref{fig:4}, with pion-loop effects added at order N$^3$LO. Four-body forces start to appear at this order as well, actually with no additional low-energy constants. 

\subsection{In-medium chiral perturbation theory}

Chiral EFT is basically a perturbative framework. In a nuclear medium, the new ``small" scale that enters in addition to three-momenta and pion mass is the Fermi momentum, $p_F$, of the nucleons. Its value at the equilibrium density of $N = Z$ nuclear matter, $n_0 = 2 p_F^3/3\pi^2 = 0.16$ fm$^{-3}$, is $p_F = 1.33$ fm$^{-1} \simeq 1.9\, m_\pi$, small compared to the chiral scale $\Lambda_\chi = 4\pi f_\pi \sim 1$ GeV. Expansions in powers of $p_F/\Lambda_\chi < 0.3$ are thus likely to converge. Even at densities $n = 3\,n_0$ the Fermi momentum still satisfies $p_F/\Lambda_\chi < 0.4$. These observations suggest indeed the applicability, within limits, of perturbative approaches to the nuclear many-body problem. 

Efforts to understand the properties of nuclear matter from ChEFT generally fall
into two categories. In one approach free-space two- and three-body nuclear potentials, with low-energy
constants fitted to NN scattering phase shifts and properties of bound two- and three-body systems,
are combined with a many-body method of choice to compute the energy per particle. Input parameters representing unresolved short-distance dynamics are fixed at the few-body level. In the second approach, {\it in-medium chiral perturbation theory}, the energy per particle is constructed as a diagrammatic expansion in the
number of loops involving explicit pion-exchange processes. There are two small scales, $p_F$
and the pion mass. The nuclear matter equation-of-state is given by an expansion in powers of the Fermi momentum.
The expansion coefficients are non-trivial functions of the dimensionless ratio, $p_F/m_\pi$, of the
two relevant low-energy scales in the problem. 

The new element in nuclear many-body calculations (compared to
scattering processes in the vacuum) is the in-medium nucleon propagator. For a
nucleon with four-momentum $p^\mu =(p_0, \boldsymbol{p})$ it reads:
\begin{equation}
{\cal G}(p) = (\gamma_\mu p^\mu+M_N)\left[ {i \over p^2 - M_N^2 + i \epsilon} -
2\pi \delta(p^2 - M_N^2) \,\theta(p_0) \theta(p_F -| \boldsymbol{p}|)\right]\, .
\label{imp}
\end{equation}
In the non-relativistic limit relevant for most nuclear physics applications, ${\cal G}(p)$ turns into 
\begin{equation}
G(p_0, \boldsymbol{p}) = { i \over p_0 - \boldsymbol{p}^{2}/2M_N + i \epsilon} -
2\pi \delta(p_0 - \boldsymbol{p}^{2}/2M_N) \, \theta(p_F -| \boldsymbol{p}|)\, .
\label{imp}
\end{equation}
The first term is the free-space propagator, while the second term (the medium insertion) accounts for the filled
Fermi sea of nucleons. This expression for $G(p_0, \boldsymbol{p})$ can be rewritten as a sum of
particle and hole propagators, a form more commonly used in non-relativistic many-body perturbation
theory. With the decomposition in Eq.\,(\ref{imp}), closed
multi-loop diagrams representing the energy density at zero temperature can be organized
systematically in the number of medium insertions. Thermodynamics proceeds in an analogous way for the free energy density, with $\theta(p_F -| \boldsymbol{p}|)$ replaced by corresponding thermal distributions \cite{HKW2013}.

Various applications of in-medium chiral perturbation theory have been reported in recent years, in particular, computations of the equations-of-state for nuclear and neutron matter. Representative examples are given in the following. 

\subsubsection {Nuclear chiral thermodynamics, part I}

\begin{figure}
\centerline{\includegraphics[width=7.5cm] {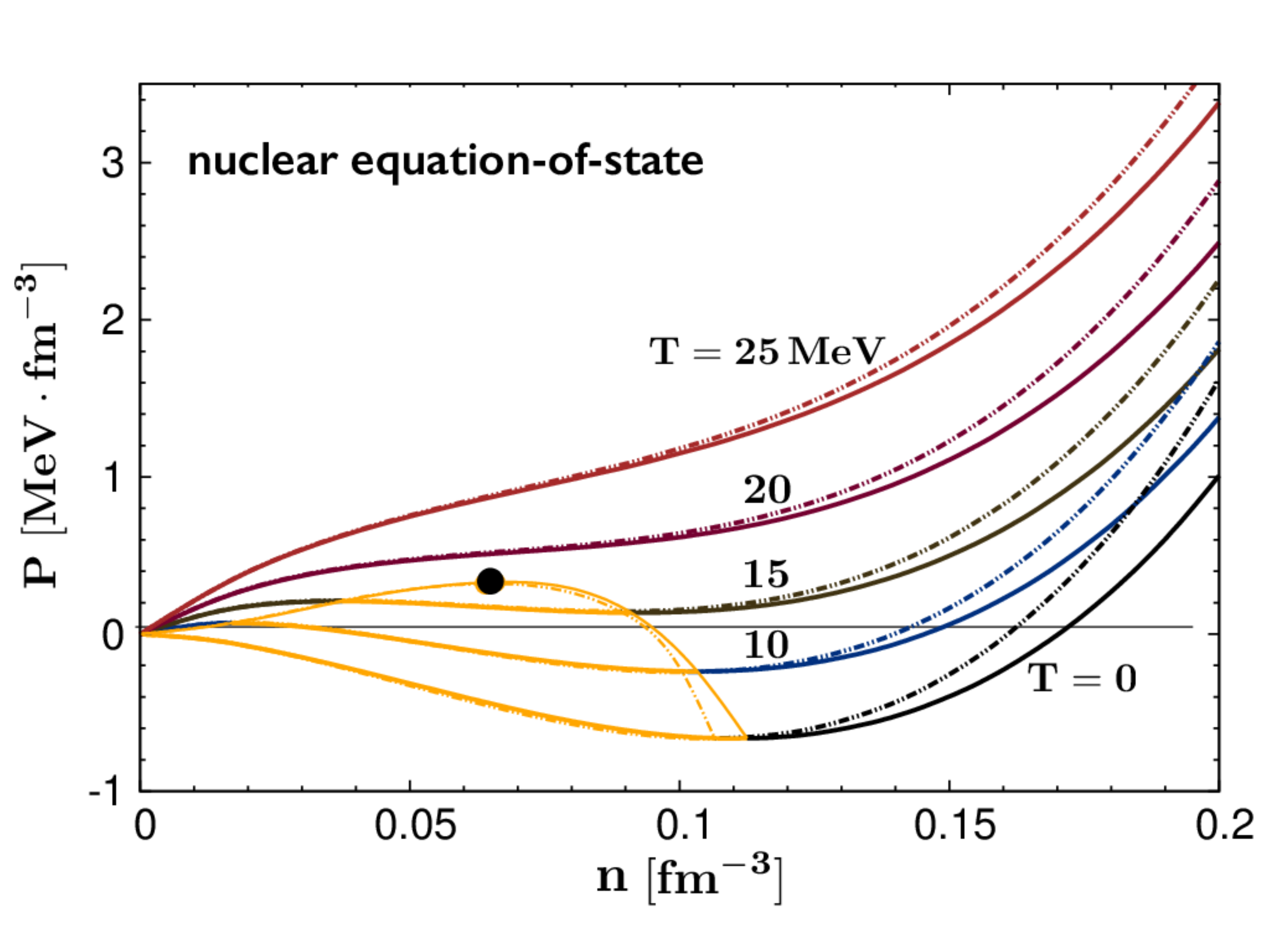}}
\caption{Equation-of-state for symmetric nuclear matter displaying the first-order liquid-gas phase transition. Isotherms of pressure as function of density are computed with $N^3LO$ chiral NN interactions and $N^2LO$ three-body forces\cite{WHKW2014,WHK2015}. Solid and dash-dotted curves correspond to cutoffs $\Lambda = 414, 450$ MeV, respectively. The calculated critical temperature is $T_c \simeq 17.4$ MeV.}
\label{fig:5}
\end{figure}

Nuclear matter at finite temperature features a liquid-gas phase transition which any realistic many-body calculation should reproduce. Calculations of the free energy of symmetric nuclear matter (with equal number of neutrons and protons, $N=Z$) have been performed using $N^3LO$ chiral NN interactions plus $N^2LO$ three-body forces in combination with Kohn-Luttinger-Ward many-body perturbation theory. The resulting equation-of-state is shown in Fig.\,\ref{fig:5}. The first-order liquid-gas phase transition is reproduced with a critical temperature of 17.4 MeV. For comparison, the critical temperature deduced from empirical data\cite{Elliot2013} is $T_c = 17.9\pm0.4$ MeV. The empirical pressure and density at the critical point, $P_c = 0.31\pm 0.07$ MeV$\cdot$fm$^{-3}$ and $n_c = 0.06\pm 0.01$ fm$^{-3}$, are also well reproduced by the ChEFT calculation as evident from Fig.\,\ref{fig:5}.

\subsubsection{Nuclear and neutron matter at zero temperature}

\begin{figure}
\centerline{\includegraphics[width=8.5cm] {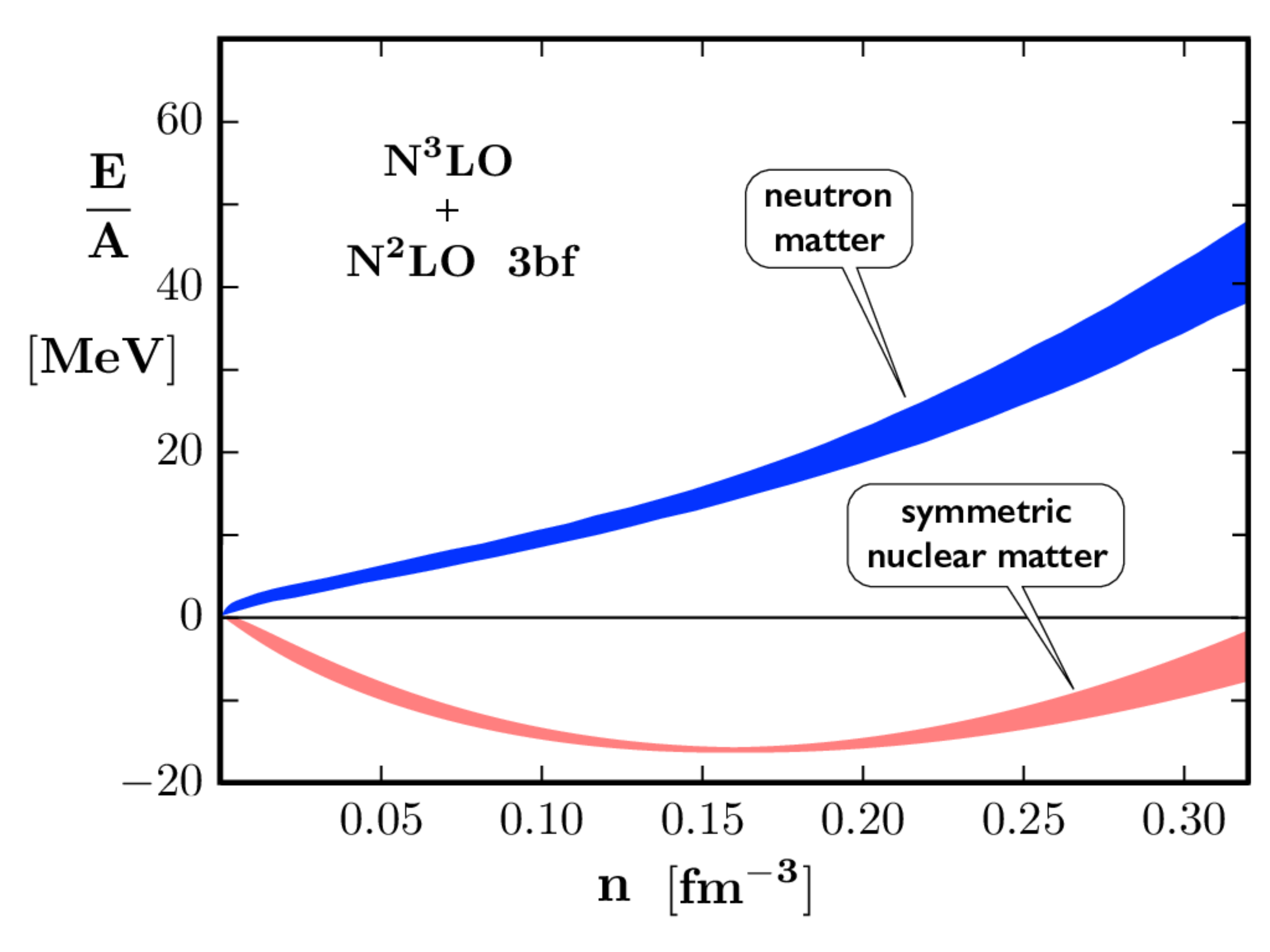}}
\caption{Energy per particle of symmetric nuclear matter and neutron matter at temperature $T=0$ as function of baryon density. Results of ChEFT calculations using with $N^3LO$ chiral NN interactions and $N^2LO$ three-body forces. Uncertainty bands correspond to cutoff range $\Lambda = 450 - 500$ MeV. Adapted from ref.\cite{LH2018} .}
\label{fig:6}
\end{figure}

Examples of recent perturbative ChEFT computations of the energy per particle, $E/A$, for nuclear and neutron matter at $T = 0$ are shown in Fig.\,\ref{fig:6}. Starting from $N^3LO$ chiral NN interactions and $N^2LO$ three-body forces, calculations using third-order many-body perturbation theory have been performed \cite{HK2017,LH2018} . These calculations are considered reliable up to densities $n \sim 2\,n_0$ and perhaps slightly beyond, but because of their perturbative nature, they cannot be extended up to the densities relevant for neutron star cores. 

Further-reaching studies have recently been performed deriving a nuclear energy-density functional (EDF) \cite{LH2017} from chiral EFT. This EDF reproduces local density distributions of finite nuclei such as $^{208}$Pb very well. When extrapolated to neutron star densities, the emerging pressure as function of energy density turns out to be sufficiently high in order to support two-solar-mass neutron stars with radii around 11-12 km. 

\subsection{Non-perturbative methods: Functional Renormalization Group}

The perturbative chiral EFT approach to nuclear and neutron matter relies on the convergence of an expansion of the energy-density in powers of $p_F/\Lambda$, the Fermi momentum over a suitably chosen momentum space cutoff. Typical ChEFT cutoffs are $\Lambda \sim 400-500$ MeV, about half of the chiral symmetry breaking scale $\Lambda_\chi = 4\pi f_\pi$.  At $n \sim 5\,n_0$ the Fermi momenta are comparable to $\Lambda$ and therefore a perturbative expansion is not expected to work. A non-perturbative strategy needs to be developed. In particular, multi-nucleon correlations grow rapidly with increasing density and must be treated through resummations. 

The focus will now be on functional renormalization group (FRG) methods applied to nuclear many-body systems. Here we give a brief outline and summary of developments and results reported in refs.\cite{DW2017,DW2015,DHKW2013} where formalisms and more details can be found. The underlying logic is the following: in the domain of spontaneously broken chiral symmetry, the active degrees of freedom are pions and nucleons. A corresponding hierarchy of scales starts at the chiral symmetry breaking scale, $\Lambda_\chi = 4\pi f_\pi$. This is taken as the ``ultraviolet" (UV) initialization of a renormalization group flow equation that describes the evolution of the action down to low-energy ``infrared" (IR) scales characteristic of the nuclear many-body problem at Fermi momenta $p_F\ll  \Lambda_\chi$. At the UV scale a chiral nucleon-meson Lagrangian based on the linear sigma model with an appropriate potential is chosen as a starting point. This chiral nucleon-meson field theory involves a scalar $\sigma$ field accompanying the pion as a chiral partner. The primary $\sigma$ is heavy, with a mass close to 1 GeV reminiscent of the $f_0(980)$. It is not to be confused with the broad ``$\sigma(500)$" which is generated dynamically as a pole in the s-wave $\pi\pi$ scattering T -matrix. 

The effective action in the IR limit then emerges by solving the non-perturbative FRG flow equations. The physics results in this long-wavelength, low-density limit should match those from perturbative chiral effective field theory. The FRG treatment of fluctuations involving pions has a correspondence (although not one-to-one) in the loop expansion of ChEFT. It is then an interesting point to compare (non-perturbative) FRG results with (perturbative) in-medium ChEFT calculations, in particular with reference to convergence issues in the latter. 
The strength of the chiral FRG approach is that it incorporates resummations to all orders of important multi-pion fluctuations, nucleon-hole excitations (i.e. fluctuations around the nuclear Fermi surface) and many-body correlations. It can therefore be extended up to high baryon densities as long as the system remains in the spontaneously broken (Nambu-Goldstone) realisation of chiral symmetry.

\begin{figure}
\centerline{\includegraphics[width=8.5cm] {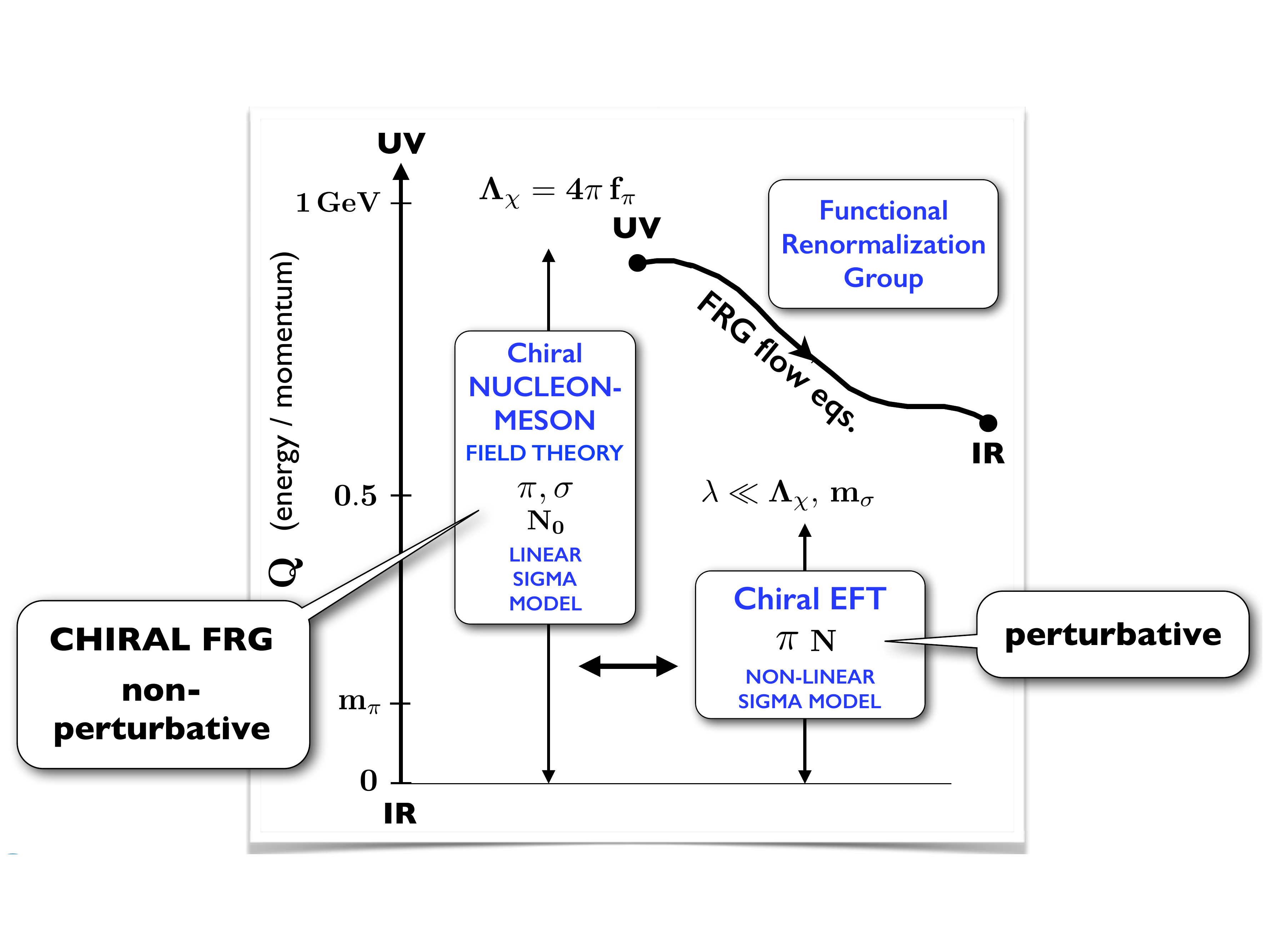}}
\caption{Illustration of renormalization group evolution concepts: from the (UV) scale of spontaneous chiral symmetry breaking in QCD, $\Lambda_\chi = 4\pi f_\pi \sim 1$ GeV, to the low-momentum (IR) scale relevant for nuclear physics.}
\label{fig:7}
\end{figure}

Fig.\,\ref{fig:7} illustrates what has just been described: a linear sigma model, treated non-perturbatively with inclusion of nucleons, undergoes RG evolution from the chiral UV scale, $\Lambda_\chi \sim 1$ GeV, all the way down to the effective action in the long-wavelength limit. At scales below about $0.5$ GeV the heavy $\sigma$ boson decouples and the theory can be rephrased in terms of a non-linear sigma model (ChEFT) with pions as the only remaining ``light" degrees of freedom, coupled to the ``heavy" nucleons. Note that the linear and non-linear sigma models are not equivalent at a perturbative level: in confrontation with observables, resummations to high orders in the linear sigma model must be perfomed when comparing to leading orders in the non-linear sigma model. But such resummations are just what the FRG equations generate. 

\subsubsection{Chiral nucleon-meson field theory} 

As mentioned, a suitable ansatz to prepare the UV input at $\Lambda_\chi \sim 1$ GeV for the FRG flow equations is a chiral field theory of mesons and nucleons, based on a linear sigma model with a non-linear effective potential. The starting Lagrangian is:
\begin{eqnarray}
{\cal L} &=& \bar{N}\left[i\gamma_\mu\partial^\mu - g(\sigma + i\gamma_5\,\boldsymbol{\tau\cdot\pi})\right]N \nonumber  \\ &+& {1\over 2}\left(\partial_\mu \sigma \partial^\mu \sigma + \partial_\mu \boldsymbol{\pi}\cdot\partial^\mu \boldsymbol{\pi}\right) - {\cal U}(\sigma, \boldsymbol{\pi})+ \Delta{\cal L}~.
\label{eq:ChNM}
\end{eqnarray}
The $\Delta{\cal L}$ term of this Lagrangian represents short-distance dynamics expressed in terms of isoscalar and isovector vector fields coupled to nucleons, corresponding to contact interactions in ChEFT. The potential ${\cal U}(\sigma, \boldsymbol{\pi})$ is written as a polynomial up to fourth order in the chiral field, $\chi \equiv \sigma^2 + \boldsymbol{\pi}^2$, plus a $\chi^2\log\chi$ term and a symmetry breaking piece proportional to $m_\pi^2\sigma$. This potential ${\cal U}$ is constructed such as to be consistent with pion-nucleon data and selected ground state properties of nuclear matter. 

The nucleon mass $M_N = g\sigma$ is coupled dynamically to the scalar field. Its expectation value, $\langle\sigma\rangle$ is normalized to the pion decay constant $f_\pi$ in the vacuum so that the mass of the free nucleon satisfies the Goldberger-Treiman relation, $M_N = g\,f_\pi$, with the axial vector constant chosen here to be $g_A =1$. In general, $\langle\sigma\rangle$ acts as an order parameter for spontaneous chiral symmetry breaking. The region of  temperatures $T$ and densities $n$ where this chiral FRG framework can be applied is defined by non-zero $\langle\sigma\rangle(T,n)\equiv f^*_\pi(T,n)$. 

Finite temperatures and chemical potentials are treated using the Matsubara formalism. Minkowski space-time is Wick-rotated to Euclidean space. Time components are transformed as $x^0\rightarrow-i\tau$. The $\tau$-dimension is compactifed on a circle, such that $\tau$ is restricted to $[0,\beta]$ with the inverse temperature $\beta=1/T$. Time-integrals are replaced by $-i\int_0^\beta d\tau$. Boson and fermion fields are periodic or anti-periodic, respectively, under $\tau\rightarrow \tau+\beta$. The Minkowski-space action $S=\int d^4x\;\mathcal L$ is replaced by the Euclidean action $S_{\rm E}=\int_0^\beta d\tau\int d^3x\;{\cal L}_{\rm E}$. 

The pertinent steps are then the following. An effective action, $\Gamma_k[\Phi]$ depending on a renormalization scale $k$, is introduced, where $\Phi$ stands for the set of all chiral boson and nucleon fields. The action derived from ${\cal L}_E$, the Euclidean version of the Lagrangian (\ref{eq:ChNM}), serves as the initialization of $\Gamma_k$ at the UV scale,  $k_{UV} \sim \Lambda_\chi = 4\pi f_\pi$. The flow of $\Gamma_k$ is determined in such a way that it interpolates between the UV action and the full quantum effective action $\Gamma_{\rm eff}=\Gamma_{k=0}$ in the infrared limit, $k\rightarrow 0$. The evolution of $\Gamma_k$ as a function of $k$ is given by Wetterich's flow equation \cite{We1993}, schematically written as 
\begin{align}\label{eq:Wetterich}
	\begin{aligned}
		k\,\frac{\partial\Gamma_k[\Phi]}{\partial k}=
		\begin{aligned}
			\includegraphics[width=0.15\textwidth]{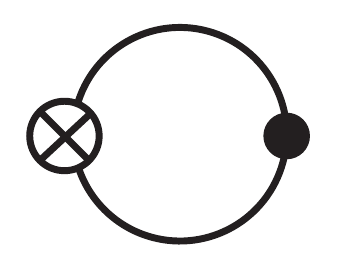}
		\end{aligned} \vspace{-1cm}=\frac 12 {\rm Tr}\left[k\frac{\partial R_k}{\partial k}\cdot\Big(\Gamma_k^{(2)}[\Phi]+R_k\Big)^{-1}\right]~~.
	\end{aligned}
\end{align}
The trace Tr stands for all relevant sums and integrations. A scale regulator, $R_k(p)$, is introduced in order to restrict momenta $p$ in loop integrals to $p^2 < k^2$. The derivative $\partial_k R_k$ has maximum weight at $p^2 \simeq k^2$. The matrix $\Gamma_k^{(2)}$ collects 2nd functional derivatives of the effective action with respect to chiral and nucleon fields. It represents the {\it full} inverse propagators of all particles involved. In the pictorial illustration of the flow equation these full propagators are marked by the dot on the loop line while the $k$-regulator is symbolized by the crossed circle.

The welcome feature of the FRG system of equations is its capability to generate fluctuations to all orders beyond mean-field approximation. The ``soft" degrees of freedom that contribute most prominently to these fluctuations  are the pion field with its small mass and low-energy nucleon-hole excitations. Both these types of excitations enter non-perturbatively through their full propagators in the flow equation (\ref{eq:Wetterich}). 

For the treatment of the thermodynamics with inclusion of fluctuations it is useful to compute the flow of the difference between the effective action at given values of temperature and chemical potential, $\Gamma_k(T,\mu)$, as compared to the potential at a reference point for which we choose equilibrium nuclear matter at zero temperature, $\Gamma_k(0,\mu_0)$ with $\mu_0 = M_N + E_0/A = 923$ MeV. The flow of the difference, $\bar\Gamma_k=\Gamma_k(T,\mu)-\Gamma_k(0,\mu_0)$, satisfies the FRG equation
\begin{align}
	\begin{aligned}
		\frac{k\,\partial\bar \Gamma_k}{\partial k}(T,\mu)&=
		\begin{aligned}
			\hspace{-.1cm}
			\vspace{1cm}
			\includegraphics[width=0.08\textwidth]{wetterich_fermion}
		\end{aligned} \vspace{-1cm}\Bigg|_{T,\mu}-
		\begin{aligned}
			\hspace{-.1cm}
			\vspace{1cm}
			\includegraphics[width=0.08\textwidth]{wetterich_fermion}
		\end{aligned} \Bigg|_{\begin{subarray}{l} T=0 \\ \mu=\mu_0 \end{subarray}}.
	\end{aligned}
\end{align}
The actual computational work involves some simplifying assumptions and approximations: the effective action is treated in leading order of the derivative expansion and we work in the local potential approximation, neglecting (small) wave function renormalization effects on the chiral boson fields  and possible higher order derivative couplings. Moreover, the $k$-running of the Yukawa coupling $g$ is ignored; the dependence of the nucleon mass on temperature and chemical potential scales with that of the pion decay constant, $f_\pi^*(T,\mu) = \langle\sigma\rangle(T,\mu)$. 

\subsubsection{Nuclear chiral thermodynamics, part II}

\begin{figure}
	\centerline{\includegraphics[width=6.5cm] {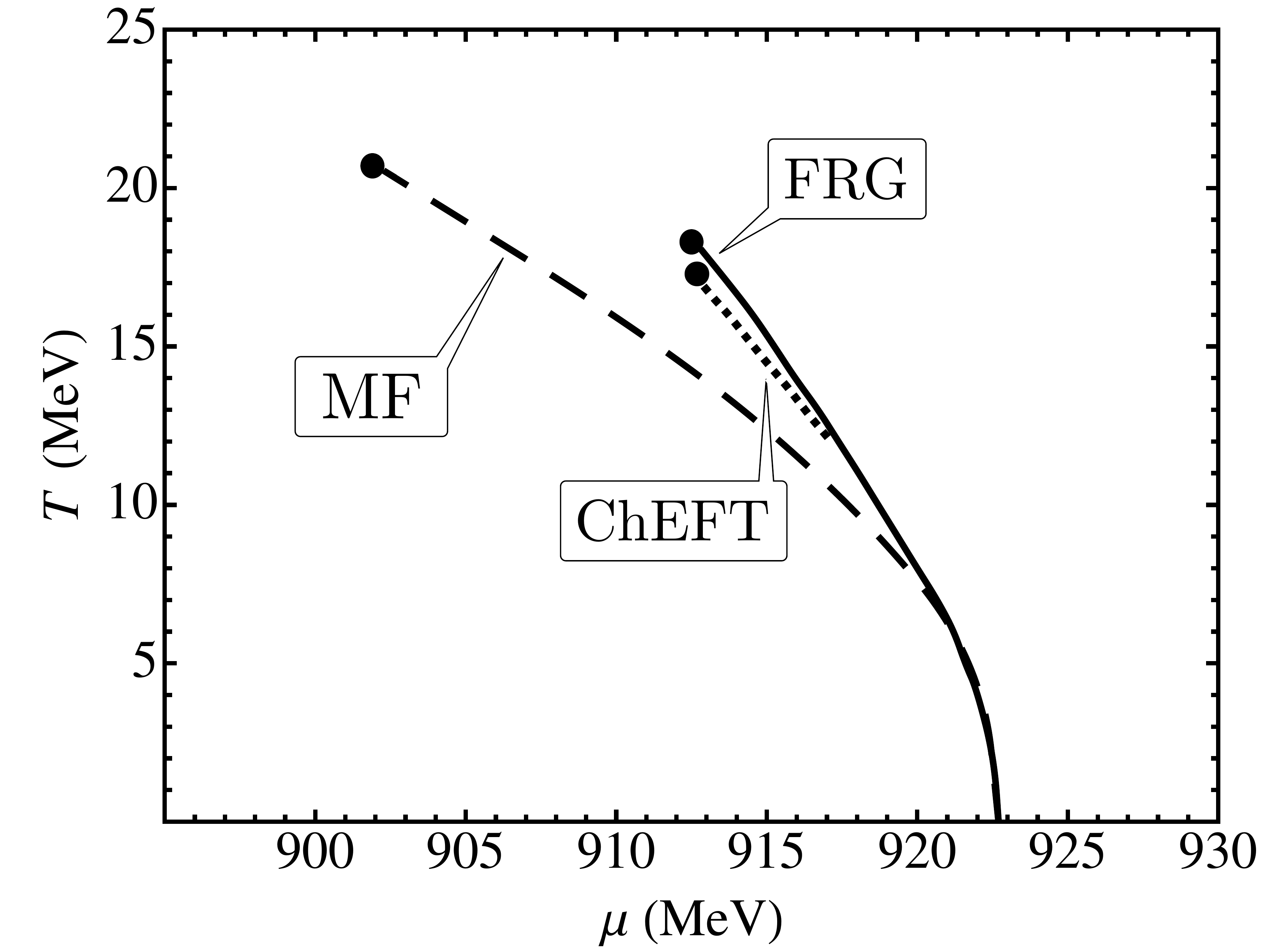}}
	\caption{Liquid-gas phase transition of symmetric nuclear matter in a $T-\mu$ diagram. Solid curve (FRG): chiral FRG calculation\cite{DW2015} including bosonic fluctuations and particle-hole excitations. Dotted curve (ChEFT): perturbative chiral EFT calculation\cite{WHKW2014} using $N^3LO$ interactions together with $N^2LO$ three-body forces. Dashed line (MF): mean-field approximation result of chiral nucleon-meson theory.}
	\label{fig:8}
\end{figure}

We proceed by examining once again the liquid-gas phase transition in symmetric nuclear matter, now using the chiral FRG approach. The $T-\mu$ phase diagram of Fig.\,\ref{fig:8} shows the first-order liquid-gas transition lines calculated using the full FRG scheme\cite{DHKW2013} of chiral nucleon-meson field theory, together with the corresponding result from a perturbative second order Kohn-Luttinger-Ward many-body calculation\cite{WHKW2014} with chiral EFT input at $N^3LO$. The comparison demonstrates the remarkably close proximity of these two results, both reproducing the empirical critical point within uncertainties. The chiral FRG result for the critical temperature is $T_c = 18.3$ MeV. Evidently, perturbative chiral EFT and the nonperturbative FRG treatment of the chiral nucleon-meson model yield very similar physics at moderate temperatures $T\lesssim 20$ MeV und densities around and below $n_0 \simeq 0.16$ fm$^{-3}$. However, the mean-field approximation of the chiral FRG effective action, replacing fields by their expectation values and thus ignoring fluctuations, gives a qualitatively different picture with a critical point outside the empirical range. It is obvious that pionic and nucleon-hole fluctuations beyond mean field are important for a realistic treatment of the nuclear many-body problem.

\begin{figure}
	 \centerline{\includegraphics[width=7cm] {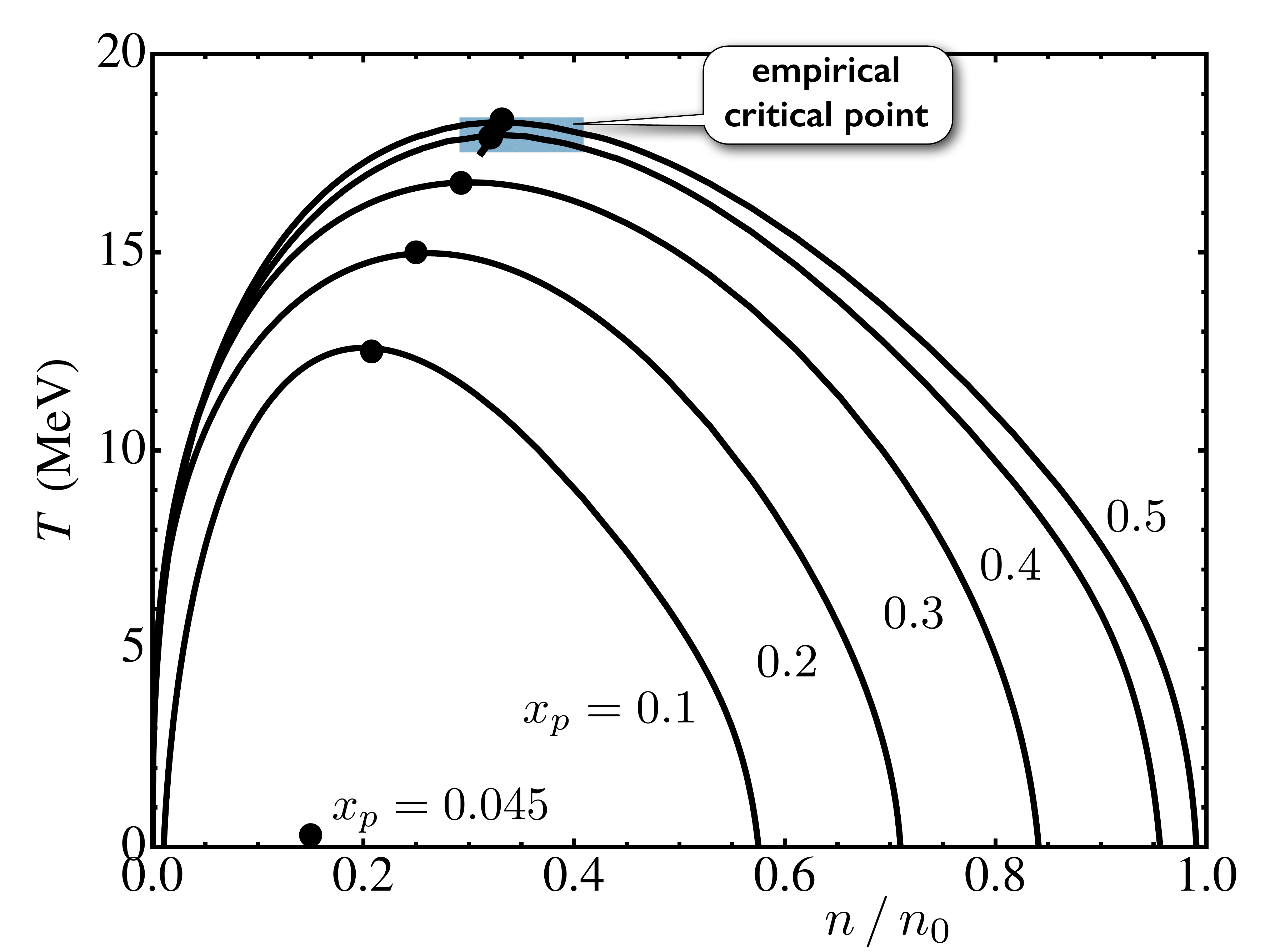}}
	\caption{The liquid-gas coexistence regions for different proton fractions $x_p = Z/A$. The trajectory of the critical point is shown as the dotted line.\label{fig:9}. The shaded rectangle indicates the empirical location of the critical point\cite{Elliot2013} for symmetric nuclear matter with $x_p = 0.5$. Figure adapted from Ref.\cite{DW2015} .}
\end{figure}

Further instructive insights can be gained by studying the liquid-gas thermodynamics of asymmetric nuclear matter, varying the proton fraction $x_p= Z/A$ systematically towards pure neutron matter. In Fig.\,\ref{fig:9} the coexistence regions in a temperature/density-plot are shown for different proton fractions. As the temperature increases, the phase coexistence region melts until it disappears at a certain $x_p$-dependent critical temperature characterized by a second-order critical endpoint. The trajectory of the critical endpoint as it evolves with decreasing proton fraction $x_p$ is indicated by the dotted curve. For $x_p$ smaller than a critical value of $x=0.045$ the energy per particle begins to increase monotonously as a function of density and the coexistence region vanishes altogether.

In summary, the non-perturbative FRG framework based on chiral nucleon-meson field theory yields results for nuclear thermodynamics that are consistent with those of perturbative chiral EFT at moderate temperatures and densities. This is true both for symmetric and asymmetric nuclear matter. In fact, with an isovector contact term fixed to reproduce a symmetry energy around 32 MeV,  the behaviour of asymmetric nuclear matter is almost entirely governed by the isospin dependence of the two- and multi-pion exchange processes between nucleons generated by chiral dynamics.
 
\subsection{Chiral symmetry restoration and order parameters}

Whereas perturbative chiral EFT has its limitations when extrapolating to densities beyond $n > 2\,n_0$, the nonperturbative chiral FRG approach can in principle be extended to compressed baryonic matter at higher densities.
A necessary condition for this to work is that matter remains in the hadronic phase chracterized by the spontaneously broken Nambu-Goldstone realisation of chiral symmetry. 

Lattice QCD computations\cite{Borsanyi2010, Bazavov2012} at $\mu = 0$ demonstrate the existence of a crossover transition towards restoration of chiral symmetry in its Wigner-Weyl realisation at temperatures $T > T_c \simeq 0.15$ GeV. Chiral symmetry is presumably also restored at large baryon chemical potentials and low temperature, although the critical value of $\mu$ at which this transition might take place is unknown. It presumably corresponds to densities that exceed $n_0$ by a large factor. 

Several calculations of isospin-symmetric matter using Nambu \& Jona-Lasinio or chiral quark-meson models have predicted a first-order chiral phase transition at vanishing temperature for quark chemical potentials, $\mu_q$, around 300\,MeV (see, e.g., Refs.\cite{Fukushima2008,RRW2007,Schaefer2007,HRCW2009,Herbst2013}). Translated into baryonic chemical potentials, $\mu\simeq3\mu_q$, chiral symmetry would then be restored not far from the equilibrium point of normal nuclear matter, $\mu_c =923$ MeV. Nuclear physics with its well-established empirical phenomenology teaches us that this can obviously not be the case. However, these calculations -- apart from the fact that they operate with (quark) degrees of freedom that are not appropriate for dealing with the hadronic phase of QCD -- work mostly within the mean-field approximation. It is therefore of great importance to examine how fluctuations beyond mean-field can change this scenario. Using chiral nucleon-meson field theory and FRG, we shall indeed point out that the mean-field approximation cannot be trusted: it is likely that fluctuations shift the chiral phase transition to extremely high baryon densities.

Within chiral nucleon-meson (ChNM) field theory the expectation value of the scalar field, $\langle\sigma\rangle$, takes over the role of the quark condensate, $\langle\bar{q}q\rangle$, as order parameter for the spontaneous breaking of chiral symmetry. In chiral EFT the temperature and density dependence of the quark condensate can be calculated by taking the derivative of the free-energy density, ${\cal F}(T,n)$, with respect to the quark mass, or equivalently, the squared pion mass. The Hellmann-Feynman theorem in combination with the Gell-Mann--Oakes--Renner relation gives the ratio of the in-medium chiral condensate to its vacuum value in the form:
\begin{align}
	\frac{\langle\bar\psi\psi\rangle(T,n)}{\langle\bar\psi\psi\rangle_0} = 1-{1\over f_\pi^2} \frac{\partial\mathcal F(T,n)}{\partial m_\pi^2}\,,
\end{align}
The corresponding ratio in the ChNM model is $\langle\sigma\rangle(T,n)/\langle\sigma\rangle_0$, with the vacuum value normalized as $\langle\sigma\rangle_0 = f_\pi$. Chiral order parameters for symmetric nuclear matter show similar trends in both ChNM-FRG and chiral EFT calculations: a mean-field approximation treatment would predict a first-order chiral phase transition for $T=0$ at a density as low as $n \simeq 1.6\,n_0$. However, once fluctuations are taken into account\cite{DHKW2013} the chiral restoration transition is shifted to chemical potentials well beyond $\mu \sim 1$ GeV (or densities $n > 3\,n_0$).

\begin{figure}
	 \centerline{\includegraphics[width=6cm] {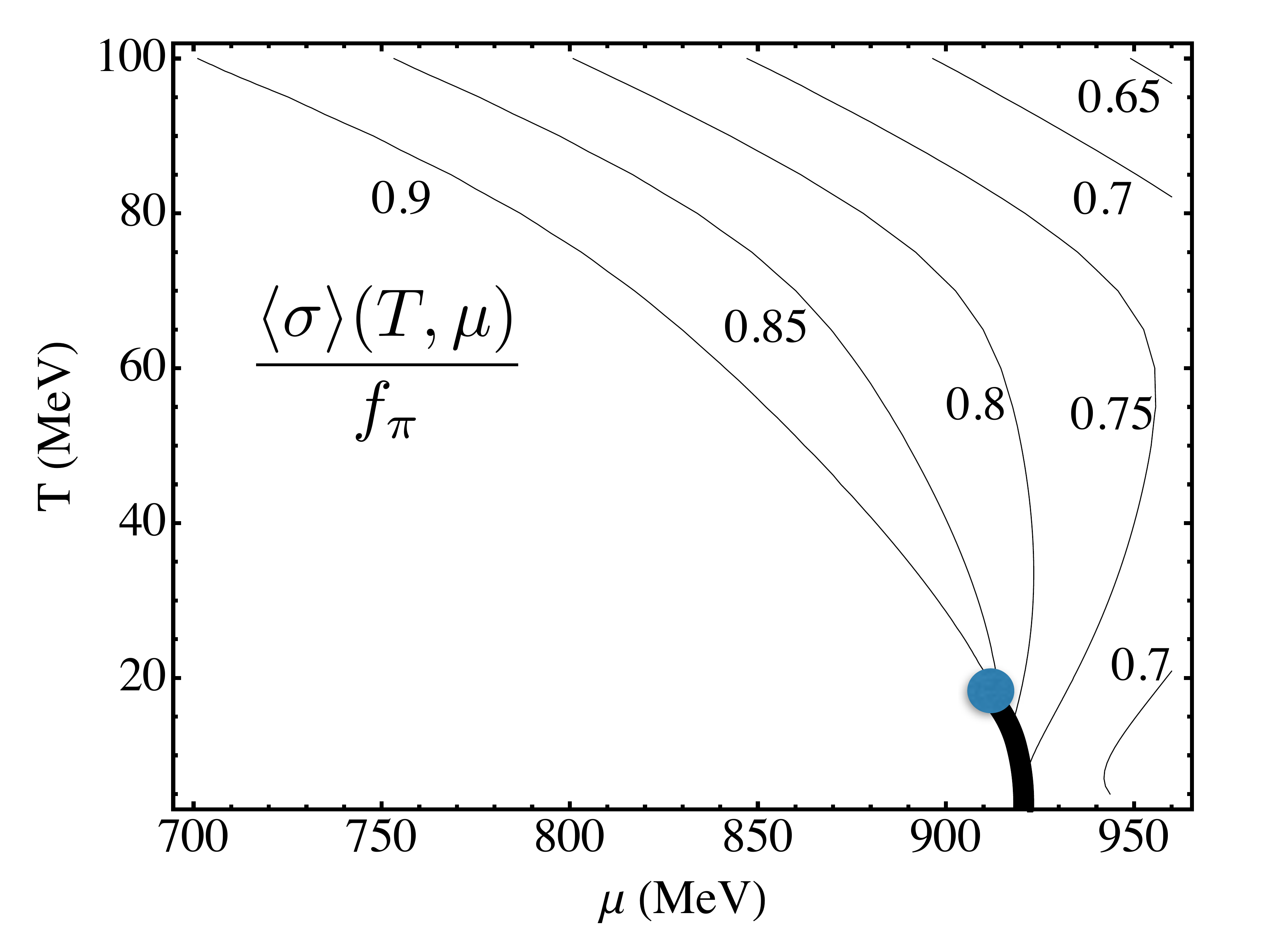}}
	\caption{$T-\mu$ phase diagram of symmetric nuclear matter calculated in the FRG-ChNM model \cite{DHKW2013}. Contour lines for constant values of the chiral order parameter $\langle\sigma\rangle/f_\pi$ are drawn with numbers attached. The liquid-gas first-order transition line and its critical point are also indicated for orientation.}
	\label{fig:10}
\end{figure}

It is instructive to examine the order parameter $\langle\sigma\rangle$ in the $T$--$\mu$ phase diagram around the liquid-gas transition. Figure\,\ref{fig:10} shows contours of constant values $\langle\sigma\rangle/f_\pi$ in the $T-\mu$ plane calculated using the chiral nucleon-meson theory in combination with full FRG including fluctuations. Evidently, within the whole region of temperatures $T\lesssim 100$ MeV and baryon chemical potentials $\mu\lesssim 1$ GeV, the order parameter remains far from zero and there is no tendency towards a chiral phase transition. Nowhere in this whole $(T,\mu)$ range does the effective nucleon mass in the medium drop below $M^*_N(T,\mu) \simeq 0.7\,M_N$. 

\begin{figure}
	 \centerline{\includegraphics[width=6cm] {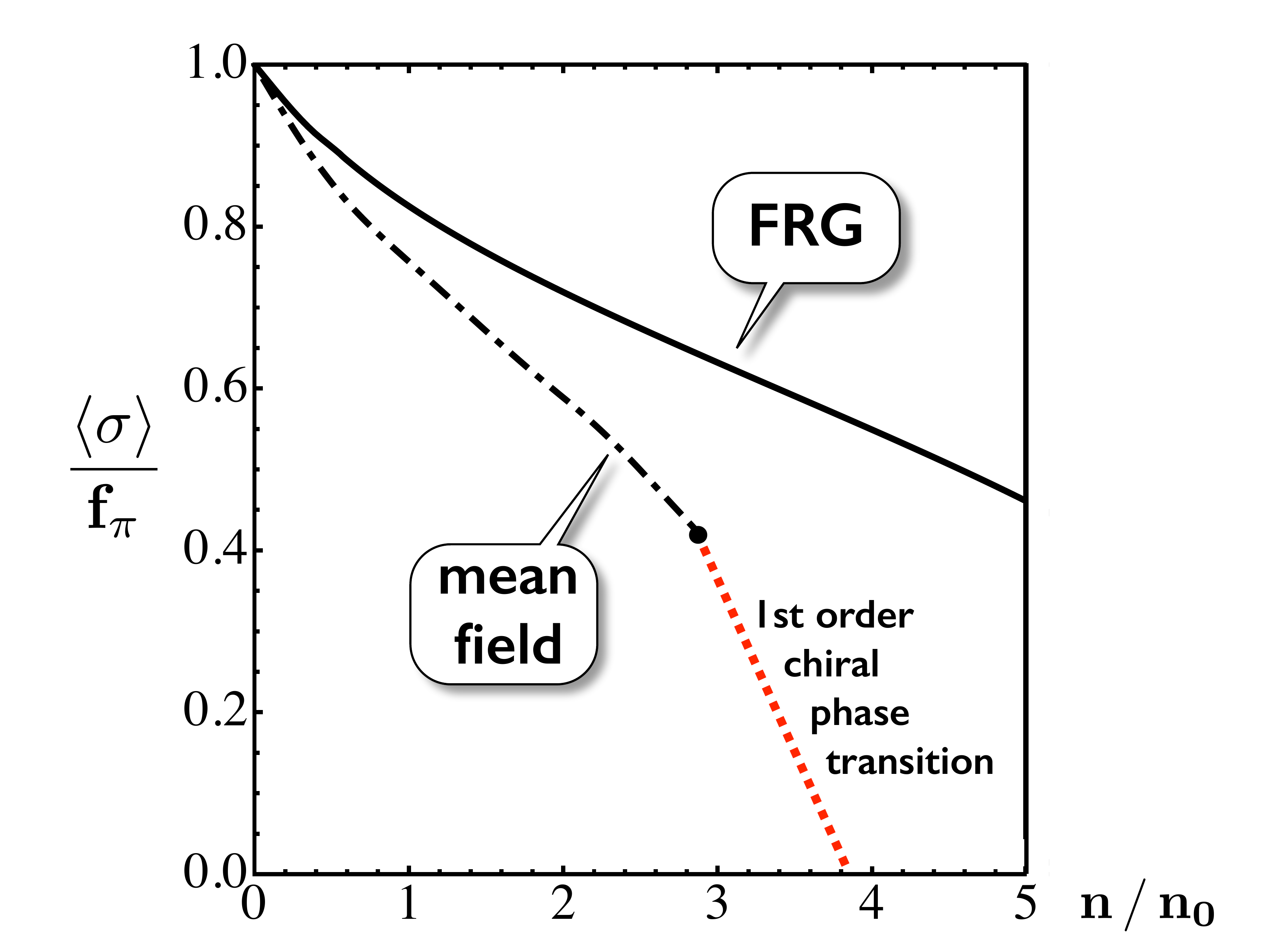}}
	\caption{Density dependence of the chiral order parameter for pure neutron matter at vanishing temperature\cite{DW2015} . Solid line (FRG): FRG calculation with chiral nucleon-meson theory input, including fluctuations. Dashed line (MF): mean-field approximation result featuring a first-order chiral phase transition.}
	\label{fig:11}
\end{figure}

For pure neutron matter at $T=0$, the chiral condensate has been calculated previously within (perturbative) chiral effective field theory\cite{Kaiser2009,Krueger2013} . Chiral nuclear forces treated up to four-body interactions\cite{Krueger2013} at N$^3$LO were shown to work moderately against the leading linearly decreasing condensate with increasing density around and beyond $n \simeq n_0$. The non-perturbative FRG approach permits an extrapolation to higher densities.  Results are presented in Fig.\,\ref{fig:11}. In mean-field approximation the order parameter $\langle\sigma\rangle/f_\pi$ shows a first-order chiral phase transition at a density of about $3\,n_0$. However, the situation changes qualitatively when fluctuations are included using the FRG framework. The chiral order parameter now turns into a continuous function of density, with no indication of a phase transition. Even at five to six times nuclear saturation density the order parameter still remains at about forty percent of its vacuum value. Only at densities as large as $n \sim 7n_0$ does the expectation value of $\sigma$ show a more rapid tendency of a crossover towards restoration of chiral symmetry. But this is even beyond the range of densities that may be reached in the cores of neutron stars. 

As a significant outcome of the nonperturbative FRG computations, we thus observe a huge influence of higher order fluctuations involving Pauli blocking effects in multiple pion-exchange processes and multi-nucleon correlations at high densities. With neutron matter remaining in a phase of spontaneously broken chiral symmetry even up to such very high densities, this encourages further-reaching applications and tests of the FRG-ChNM approach in constructing an equation of state for the interior of neutron stars. 

\section{Implications for Neutron Stars}

Neutron star observations provide stringent constraints for the equation-of-state (EoS) of highly compressed baryonic matter. New standards have been set by the discovery of heavy neutron stars\cite{Demorest2010,Antoniadis2013} with masses $M_{n-star} \simeq 2\,M_{\odot}$. Whatever the detailed composition of dense matter may be: its EoS must produce sufficiently high pressure in order to stabilize two-solar-mass neutron stars against gravitational collapse. Additional important constraints are now derived from the analysis of gravitational wave signals\cite{Abbot2017} produced by neutron star mergers.

\begin{figure}
	 \centerline{\includegraphics[width=8cm] {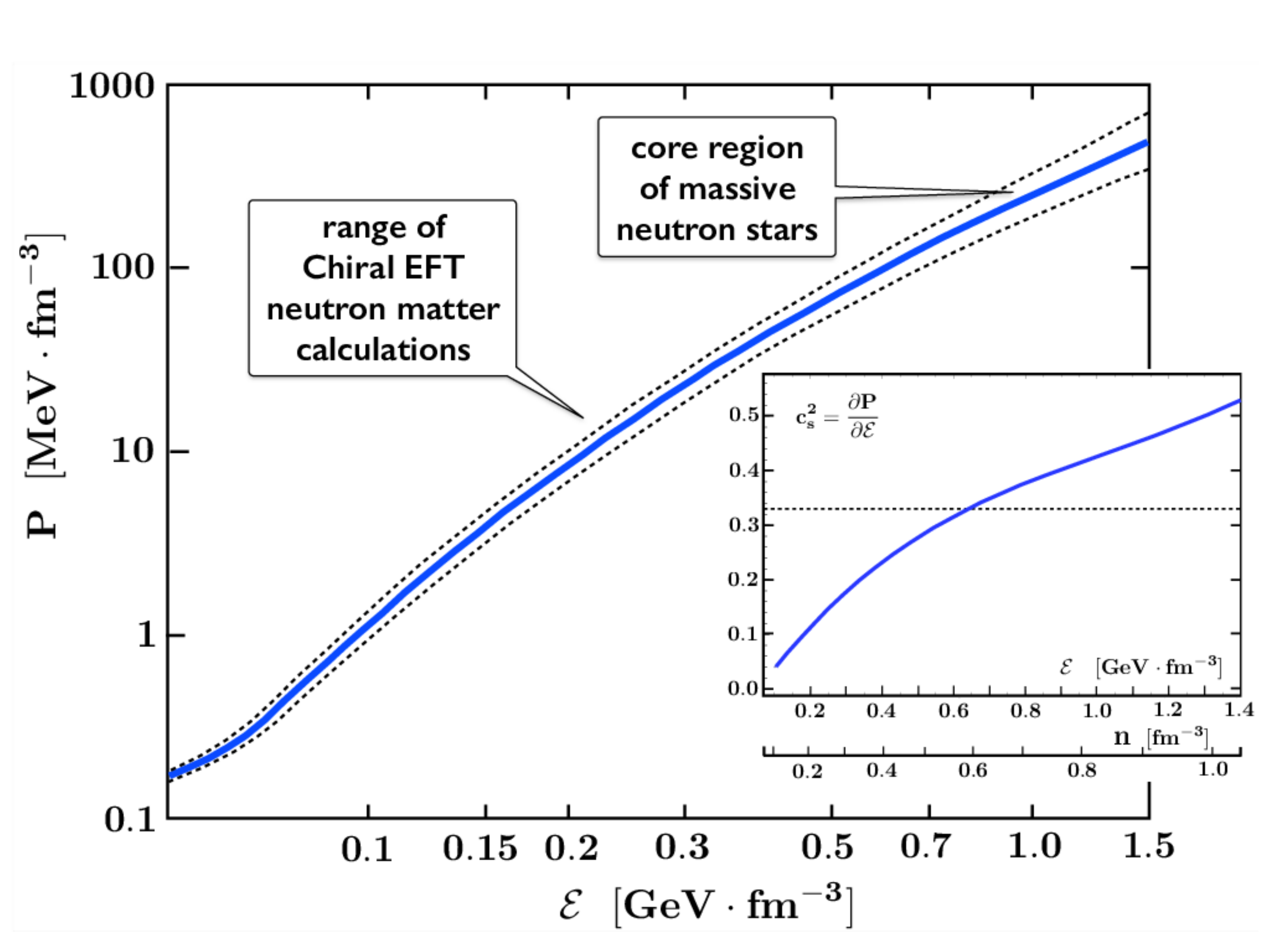}}
	\caption{Pressure as function of energy density fo neutron star matter with inclusion of beta equilibrium. Solid curve: result of a calculation\cite{DW2017,DW2015} starting from chiral nucleon-meson theory and solving functional renormalization group equations. The uncertainty band indicated by dotted lines corresponds to a range of input symmetry energy values 30 - 34 MeV at density $n = n_0 = 0.16$ fm$^{-3}$. Also shown is the square of the sound velocity, $c_s^2 = \partial P({\cal E})/\partial{\cal E}$.}
	\label{fig:12}
\end{figure}

\subsection{Equation-of-state for neutron star matter}

A realistic equation-of-state for neutron star matter at $T=0$ that turns out to satisfy the constraints just mentioned has been computed\cite{DW2015,DW2017} solving the FRG equations with input from chiral nucleon-meson theory. At low densities this EoS is consistent with chiral EFT calculations of neutron matter and symmetric nuclear matter. The symmetry energy at $n = n_0$ is $[E(Z=0) - E(Z=N)]/A = 32$ MeV. Beta equilibrium conditions for neutron star matter are properly incorporated. The result for the pressure as function of the energy density, $P({\cal E})$, in Fig.\,\ref{fig:12} shows a steep rise towards pressures exceeding 100 MeV/fm$^{3}$ in the region relevant for the core of heavy neutron stars. This amount of pressure is indeed capable of supporting a two-solar-mass neutron star. Its radius is predicted to be $R\simeq 11.5$ km. Notably the baryon density in the center of such an object does not exceed $n\sim 5\,n_0$. Following the discussion in Sec. 1.4, it then appears justified to work with nucleons and pion fields as relevant degrees of freedom even at such extreme but not hyperdense conditions.

A further interesting property of compressed baryonic matter is its velocity of sound. For a non-interacting relativistic Fermi gas the squared sound velocity has a canonical value,
\begin{equation}
c_s^2= {\partial P({\cal E})\over \partial{\cal E}} = {1\over 3}~,
\end{equation}
which is supposed to be reached at asymptotically high densities. The inset of Fig.\,\ref{fig:12} shows that the squared sound velocity of the FRG - ChNM equation-of-state exceeds $c_s^2 = 1/3$ at a baryon density around $n\sim 4\,n_0$ and continues to grow as it approaches neutron star core densities. This behaviour can be traced to the continuously rising strength of repulsive many-body correlations driven in part by the Pauli principle as the density increases. At much higher densities, once nucleon clusters dissolve and quark matter takes over, $c_s^2$ is expected to decrease again and ultimately approach its asymptotic value of 1/3 from below at ultrahigh densities\cite{TCGR2018} . 

\begin{figure}
	 \centerline{\includegraphics[width=8cm] {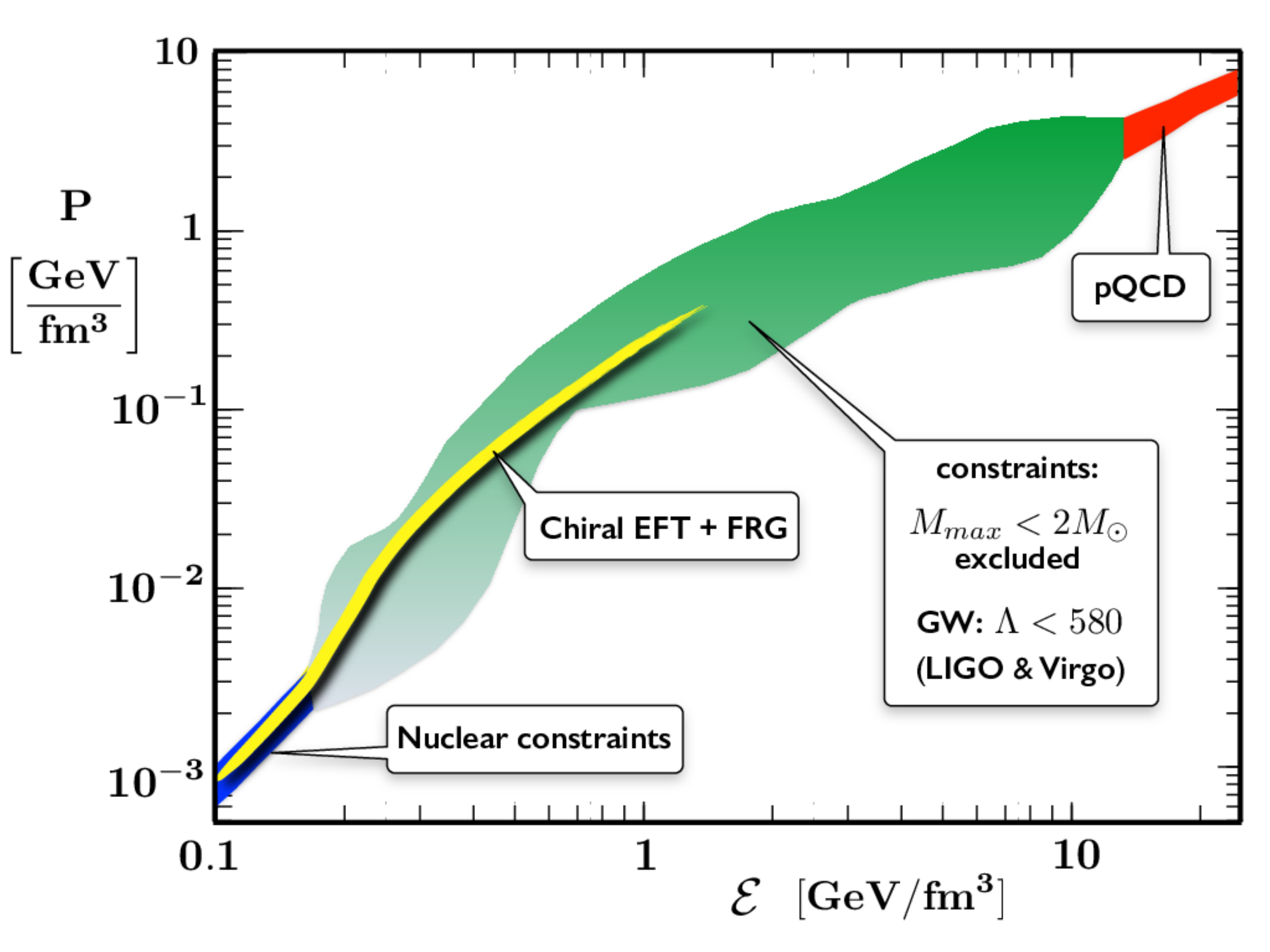}}
	\caption{Pressure as function of energy density for neutron star matter with constraints from perturbative QCD and nuclear EFT calculations at the upper and lower ends of the density scale, adapted from refs.\cite{Annala2018,Vuorinen2018}. The shaded area represents the range of acceptable equations-of-state subject to empirical conditions on neutron star maximum mass and tidal deformability, the latter from LIGO \& Virgo gravitational wave analysis. The curve denoted ``Chiral EFT + FRG" represents $P({\cal E})$ from Fig.\,\ref{fig:12}. }
	\label{fig:13}
\end{figure}

An important effort presently pursued and steadily being improved is to constrain the neutron star equation-of-state systematically from observational data, together with an interpolation between theoretical limits provided by nuclear physics at low densities and perturbative QCD at extremely high densities. Fig.\,\ref{fig:13} shows a recent example. Nuclear constraints as represented by ChEFT calculations\cite{HKW2013,HLPS2013} set the low-density limit of $P({\cal E})$ at energy densities ${\cal E}\lesssim 200$ MeV/fm$^3$. Sophisticated perturbative QCD calculations\cite{Kurkela2014} determine the pressure at extreme energy densities, ${\cal E} > 10$ GeV/fm$^3$. Constraints in the region between these extremes are introduced by studying large sets of parametrized equations-of-state subject to the condition that they all produce a maximum neutron star mass of at least $2\,M_\odot$ and at the same time fulfill the tidal deformability limit\cite{Abbot2017} $\Lambda_t < 580$ deduced from the updated LIGO \& Virgo gravitaional wave analysis. The shaded area in Fig.\,\ref{fig:13} defines the region of acceptable neutron star equations-of-state under such conditions\cite{Annala2018, Vuorinen2018} . Remarkably, the EoS computed using the FRG-ChNM approach and shown in Fig.\,\ref{fig:12} satisfies these constraints up to the densities relevant for the core of massive neutron stars. Of course, one order of magnitude in the pressure $P$ still separates this neutron star domain from the perturbative QCD sector. It is in the range of densities $n > 5\,n_0$ where one can expect a (probably continuous) hadrons-to-quarks transition\cite{Baym2018} to take place.

\subsection{Outlook: strangeness in neutron stars ?}

An issue that still needs to be resolved is the so-called ``hyperon puzzle" in neutron stars. At densities around 2-3 times $n_0$ the neutron Fermi energy reaches a point at which it begins to be energetically favourable replacing neutrons by $\Lambda$ hyperons, as long as only $\Lambda N$ two-body forces are employed\cite{DSW2010,LLGP2015} . Then, however, the EoS becomes too soft and misses by far the $2\,M_\odot$ constraint for the neutron star masses. 

Ongoing investigations suggest a possible way of resolving this puzzle. The starting point is an extension to chiral $SU(3)\times SU(3)$ meson-baryon EFT and the construction of interactions now involving the complete baryon and pseudoscalar meson octets, incorporating $SU(3)$ breaking effects through the physical mass differences within the multiplets.
Hyperon-nucleon interactions constructed in this scheme\cite{Haidenbauer2013} at NLO\footnote{The statistical quality of the existing empirical data base for hyperon-nucleon scattering is still too limited to warrant more detailed studies beyond NLO.} indicate strong $\Lambda N \rightarrow \Sigma N$ coupled-channels effects in combination with repulsive short-distance dynamics which work to raise the onset condition for the $\Lambda$ chemical potential, $\mu_\Lambda = \mu_n$, towards higher densities.  Perhaps more significantly, $\Lambda NN$ three-body forces\cite{Petschauer2016,Petschauer2017} as they emerge from chiral $SU(3)$ EFT introduce additional repulsion\cite{HMKW2017} that raises $\mu_\Lambda$ further with increasing density. Detailed studies are now performed combining these repulsive effects in order to explore whether the condition $\mu_\Lambda = \mu_n$ can still be met in neutron stars. 

\section{Concluding Remarks}

The focus in this presentation has been on guiding principles leading from QCD symmetries and symmetry breaking patterns to strongly interacting complex systems such as nuclei and dense baryonic matter. It indeed turns out that chiral symmetry and its spontaneous breakdown in conjunction with the confining and scale-invariance breaking QCD forces provide the basis for constructing effective field theories of low-energy QCD that lead a long way towards the understanding of nuclear forces, nuclear many-body systems and even baryonic matter under more extreme conditions. 

In this context, a significant outcome from a non-perturbative framework using functional renormalization group methods concerns the appearance of phase transitions in the equation-of-state of baryonic matter. While the empirically established first-order liquid-gas transition in nuclear matter is well reproduced, strong fluctuations beyond mean-field approximation prevent a first-order chiral phase transition from appearing at densities as high as those encountered in the core of neutron stars. In such a scheme the quest for the emergence of quark-hadron continuity and the transition to freely floating quarks in cold and compressed baryonic matter is passed over to even more extreme density scales.

\section*{Acknowledgements}

This review has been written with a feeling of deep gratitude and lively memories of numerous conversations and discussions that the author enjoyed with the late Professor Ernest M. Henley on various occasions and over several decades, in Seattle, Munich and at other places. Ernest will be remembered as a great colleague and friend. 

This Work is supported in part by the DFG Cluster of Excellence ``Origin and Structure of the Universe" and by the ExtreMe Matter Institute (EMMI) at the GSI Helmholtz Centre for Heavy-Ion Research, 64291 Darmstadt, Germany. The author, as an EMMI Visiting Professor 2018-19, gratefully acknowledges the hospitality at this centre.

\end{document}